\documentclass[final,5p,times,twocolumn]{elsarticle}

\usepackage[T1]{fontenc}
\usepackage[utf8]{inputenc}

\usepackage{amsmath}
\usepackage{amsfonts}
\usepackage{amssymb}
\usepackage{amsxtra}
\usepackage{array}
\usepackage{graphicx}
\usepackage{hepunits}
\usepackage{units}
\usepackage{color}
\usepackage{xspace}

\definecolor{purple}{rgb}{0.5,0,0.5}
\definecolor{blue}{rgb}{0.0,0,0.9}
\definecolor{prdblue}{rgb}{0.133,0.118,0.498}
\usepackage[colorlinks=true, pdfstartview=FitV, linkcolor=prdblue, citecolor= prdblue, urlcolor=prdblue]{hyperref}

\usepackage[mathscr,scaled=1.15]{urwchancal}
\DeclareFontFamily{OT1}{pzc}{}
\DeclareFontShape{OT1}{pzc}{m}{it}%
{<-> s * [1.15] pzcmi7t}{}
\DeclareMathAlphabet{\mathpzc}{OT1}{pzc}{m}{it}

\newcommand{\be}{\begin{equation}}
\newcommand{\bea}{\begin{eqnarray}}
\newcommand{\ee}{\end{equation}}
\newcommand{\eea}{\end{eqnarray}}

\def\1eq#1{Eq.~(\ref{#1})}

\def\2eqs#1#2{Eqs.~(\ref{#1}) and~(\ref{#2})}
\def\3eqs#1#2#3{Eqs.~(\ref{#1}),~(\ref{#2}) and~(\ref{#3})}

\biboptions{sort&compress}

\journal{Physics Letters B}

\begin{document}

\begin{frontmatter}

\title{$\,$\\[-7ex]\hspace*{\fill}{\normalsize{\sf\emph{Preprint no}. NJU-INP 032/21}}\\[1ex]
Measures of pion and kaon structure from generalised parton distributions}

\author[NJNU]{J.-L. Zhang}

\author[NKU,ICNM]{K. Raya}

\author[NKU]{L. Chang}

\author[NJU,INP]{Z.-F. Cui}

\author[UHe]{J. M. Morgado}

\author[NJU,INP]{C. D. Roberts}

\author[UHe]{J. Rodr\'{\i}guez-Quintero}

%
\address[NJNU]{
Department of Physics, Nanjing Normal University, Nanjing, Jiangsu 210023, China}
\address[NKU]{
School of Physics, Nankai University, Tianjin 300071, China}
\address[ICNM]{Instituto de Ciencias Nucleares, Universidad Nacional Aut{\'o}noma de M{\'e}xico, Apartado Postal 70-543, C.\,P.\ 04510, CDMX, M{\'e}xico}
\address[NJU]{
School of Physics, Nanjing University, Nanjing, Jiangsu 210093, China}
\address[INP]{
Institute for Nonperturbative Physics, Nanjing University, Nanjing, Jiangsu 210093, China}
\address[UHe]{
Department of Integrated Sciences and Center for Advanced Studies in Physics, Mathematics and Computation, 
University of Huelva, E-21071 Huelva, Spain\\[1ex]
Email addresses:
\href{mailto:leichang@nankai.edu.cn}{leichang@nankai.edu.cn} (L. Chang);
\href{mailto:cdroberts@nju.edu.cn}{cdroberts@nju.edu.cn} (C. D. Roberts);
\href{mailto:jose.rodriguez@dfaie.uhu.es}{jose.rodriguez@dfaie.uhu.es} (J. Rodríguez-Quintero)
}

\begin{abstract}
Pion and kaon structural properties provide insights into the emergence of mass within the Standard Model and attendant modulations by the Higgs boson.  Novel expressions of these effects, in impact parameter space and in mass and pressure profiles, are exposed via $\pi$ and $K$ generalised parton distributions, built using the overlap representation from light-front wave functions constrained by one-dimensional valence distribution functions that describe available data.  Notably, \emph{e.g}.\ $K$ pressure profiles are spatially more compact than $\pi$ profiles and both achieve near-core pressures of similar magnitude to that found in neutron stars.
%
\end{abstract}

\begin{keyword}
continuum Schwinger function methods \sep
emergence of mass \sep
Nambu-Goldstone modes -- pions and kaons \sep
nonperturbative quantum field theory \sep
parton distributions \sep
strong interactions in the standard model of particle physics
\end{keyword}

\end{frontmatter}

\noindent\textbf{1.$\;$Introduction}.
High priority is being given to experiments that can yield data interpretable in terms of generalised parton distributions (GPDs) \cite{Burkardt:2002hr, Diehl:2003ny, Burkardt:2008jw} or transverse momentum dependent parton distributions \cite{Burkardt:2008jw, Barone:2001sp, Meissner:2007rx, Barone:2010zz} and thereby used to draw three-dimensional images of hadrons.  It is thus crucial to develop methods that enable calculation of these distributions and their reliable connection to quantum chromodynamics (QCD).  Focusing on GPDs, which can yield information about the spatial distributions of mass, momentum, spin, \emph{etc}.\ within hadrons, much theoretical work and phenomenology has focused on the proton \cite{Guidal:2013rya, Anselmino:2016TR, Stefanis:2016dhq}.  Less has been devoted to mesons \cite{Stefanis:2016dhq, Mezrag:2016hnp}.

A wide-ranging elucidation of $\pi$ and $K$ structural measures is important for many reasons.  Most significantly because the internal structure of pions and kaons is far more complex than often imagined and their properties provide the clearest windows onto the phenomenon of emergent hadron mass (EHM) and its modulation by Higgs-boson interactions \cite{Horn:2016rip, Roberts:2020udq, Roberts:2020hiw, Roberts:2021xnz, Roberts:2021ppnp}.  Theory efforts in this area are especially meaningful today because new experimental facilities promise to deliver high precision data on kinematic domains that have never before been explored or have remained untouched for more than thirty years \cite{Petrov:2011pg, Denisov:2018unjF, Aguilar:2019teb, Chen:2020ijn, Horn:2020ces}.

Regarding pion GPDs, notable analyses exist, using continuum and lattice methods, \emph{e.g}.\ Refs.\,
\cite{Theussl:2002xp, Dalley:2003sz, Broniowski:2007si, Mezrag:2014jka, Fanelli:2016aqc, Kumano:2017lhr, Chen:2019lcm, Ma:2019agv, Zhang:2020ecj, Lin:2020rxa}.  For the kaon, on the other hand, foundations are just being laid, \emph{e.g}.\ Refs.\,\cite{Nam:2011yw, Xu:2018eii, Kock:2020frx}.  Thus, herein, beginning with one-dimensional valence distribution functions (DFs) that describe existing data \cite{Cui:2020dlm, Cui:2020tdf}, we develop $\pi$ and $K$ light-front wave functions (LFWFs) and, subsequently, associated GPDs via the overlap representation.  These GPDs enable the calculation of numerous $\pi$ and $K$ observables, comparisons between which highlight the impact of Higgs-boson couplings into QCD on the properties of pseudoscalar mesons, whose basic character is fixed by EHM.

\smallskip

\noindent\textbf{2.$\;$Meson Light-Front Wave Functions}.
The leading-twist LFWF for the $u$ quark in a ${\mathsf P}=u\bar h$ pseudoscalar meson may be written:
\begin{align}
\label{EqLFWFG}
%
\psi_{\mathsf P}^u(x,k_\perp^2;\zeta) & =
{\rm tr}_{\rm CD} Z_2 \int \frac{dk_3 dk_4}{\pi}
\delta(n\cdot k - x n\cdot P_{\mathsf P}) \nonumber \\
& \qquad \times \gamma_5 \gamma\cdot n \chi_{\mathsf P}(k-P_{\mathsf P}/2,P_{\mathsf P};\zeta)\,,
\end{align}
where
%
$k=(k_1,k_2,k_3,k_4)$,
$x$ is the light-front fraction of the meson's total momentum, $P_{\mathsf P}$, carried by the $u$ quark and $k_\perp=(k_1,k_2)$ is the quark's momentum in the light-front transverse plane;
the trace is over colour and spinor indices;
$Z_2$ is the quark wave function renormalisation constant;
%
$n$ is a light-like four-vector, $n^2=0$, $n\cdot P_{\mathsf P} = -m_{\mathsf P}$ in the meson rest frame, with $m_{\mathsf P}$ the meson's mass;
$\chi_{\mathsf P}$ is the meson's Poincar\'e-covariant Bethe-Salpeter wave function;
and $\zeta$ is the renormalisation scale.
The LFWF is invariant under light-front-longitudinal Lorentz boosts; so when solving bound-state scattering problems using a light-front formulation, compressed or contracted objects are not encountered \cite{1835233}, \emph{e.g}.\ the cross-section for the meson+proton Drell-Yan process is the same whether the proton is at rest or moving.

Following Ref.\,\cite{tHooft:1974pnl}, Eq.\,\eqref{EqLFWFG} defines the LFWF via light-front projection of the meson's Poincar\'e-covariant Bethe-Salpeter wave function.  This scheme is also efficacious for QCD \cite{Chang:2013pq}.  It yields a LFWF expressed in a quasiparticle basis defined by the choice of renormalisation scale.  At the hadron scale, $\zeta=\zeta_{\mathpzc H}$, the quark$+$antiquark quasiparticle pair express all properties of ${\mathsf P}$ \cite{Cui:2019dwv}; for instance, they carry all the meson's light-front momentum.  This is a feature of continuum formulations of the meson bound-state problem, \emph{e.g}.\ Refs.\,\cite{Dorokhov:2000gu, RuizArriola:2004ui, Lan:2019rba, Han:2020vjp}.  With a LFWF in hand, one has access to an array of key hadron structure measures, like form factors, distribution amplitudes (DAs) and DFs.

Practicable, realistic kernels for those equations relevant to the continuum meson bound-state problem are being developed \cite{Eichmann:2016yit, Fischer:2018sdj, Qin:2020rad, Qin:2020jig}, with results available for spectra, DAs, DFs, and form factors \cite{Shi:2015esa, Ding:2015rkn, Li:2016mah, Raya:2016yuj, Gao:2017mmp, Chen:2018rwz, Ding:2018xwy, Binosi:2018rht, Cui:2020dlm, Cui:2020tdf}.  Pointwise reliable projection of computed Bethe-Salpeter wave functions onto the light-front requires use of perturbation theory integral representations (PTIRs) \cite{Nakanishi:1969ph} for the Schwinger functions involved.  Their construction is time consuming and case specific.  Herein, therefore, we follow a different and, for the present, more insightful path.

Capitalising on analyses using continuum and lattice methods, which have delivered coherent predictions for meson DAs and DFs \cite{Roberts:2021ppnp}, we work backwards to build LFWFs that reproduce such mutually consistent results for the pion and kaon.  To that end, recall that the leading-twist meson DAs are related to the LFWF as follows \cite{Brodsky:1989pv}:
\begin{equation}
\label{eq:DA}
f_{\mathsf P} \varphi^{u}_{\mathsf P}(x,\zeta_{\mathpzc H})
=  \int \frac{dk^2_\perp}{16 \pi^3} \psi_{{\mathsf P}}^{u}\left(x,k_\perp^2;\zeta_{\mathpzc H} \right)\,,
\end{equation}
where $f_{\mathsf P}$ is the meson's leptonic decay constant.
Written in this form, the DA is unit normalised.  The DA associated with the $\bar{h}$ quasiparticle is $\varphi_{\mathsf P}^{\bar h}(x;\zeta_{\mathpzc H}) = \varphi_{\mathsf P}^u(1-x;\zeta_{\mathpzc H})$.  Notice further that the valence quark DFs may also be defined via the LFWFs \cite{Brodsky:1989pv, Cui:2020dlm, Cui:2020tdf}:
\begin{equation}\label{eq:uMzetaH}
{\mathpzc u}^{\mathsf P}(x;\zeta_{\mathpzc H}) = \int \frac{d^2k_\perp}{16\pi^3} \left| \psi_{{\mathsf P}}^{u}\left(x,k_\perp^2;\zeta_{\mathpzc H} \right) \right|^2  \,,
\end{equation}
with $\overline{\mathpzc h}^{\mathsf P}(x;\zeta_{\mathpzc H})={\mathpzc u}^{\mathsf P}(1-x;\zeta_{\mathpzc H})$.  Eqs.\,\eqref{eq:DA}, \eqref{eq:uMzetaH} can be used effectively to constrain \emph{Ans\"atze} for $\pi$ and $K$ LFWFs.
%


\begin{table}[t!]
\begin{tabular}{l|c|c|c|c|c|c|c}\hline
${\mathsf P}$ & $ m_{\mathsf P}$ & $M_u$ & $M_h$ & $\Lambda_{\mathsf P}$ & $b_0^{\mathsf P}$ & $\omega_0^{\mathsf P}$ & $v_{\mathsf P}$  \\
\hline
$\pi$ & $0.14$ & 0.31 & $\phantom{1.2} M_u$ & $\phantom{3} M_u$ & $0.275$ & $1.23\phantom{5}$ & 0\phantom{.41} \\
\hline
$K$ &  $0.49$ & 0.31 & $1.2 M_u$  &$3 M_s$ &  $0.1\phantom{75}$ & $0.625$ & $0.41$ \\\hline
\end{tabular}
\caption{
\label{tab:params}
Used in Eqs.\,\eqref{tab:params} -- \eqref{eq:spectralw}, one reproduces the pion and kaon DAs described in Refs.\,\cite{Cui:2020dlm, Cui:2020tdf} and empirical values for the meson decay constants \cite{Zyla:2020zbs}: $f_\pi=0.092\,$GeV, $f_K = 0.11\,$GeV.  (Mass dimensioned quantities in GeV.)
}
\end{table}

\smallskip

\noindent\textbf{3.$\;$Spectral Representation of Light-Front Wave Functions}.
%
Exploiting the analysis in Ref.\,\cite{Xu:2018eii}, which used PTIRs to explore the character of parton quasidistributions for mesons, we write the LFWF in the following form:
\begin{equation}
\label{psiLFWF}
\psi_{{\mathsf P}}^{u}(x,k_\perp^2;\zeta_{\mathpzc H}) =
12 \left[ M_u (1-x) + M_{\bar h} x \right] {\mathpzc X}_{\mathsf P}(x;\sigma_\perp^2) \,,
\end{equation}
where:
$M_{u,\bar h}$ are dressed-quark/-antiquark mass-scales, which should be chosen commensurate with the infrared values of the relevant running masses \cite[Fig.\,2.5]{Roberts:2021ppnp};
$\sigma_\perp=k_\perp^2+\Omega_{\mathsf P}^2$,
\begin{subequations}
\begin{align}
\nonumber
\Omega_{\mathsf P}^2  = v M_u^2 & + (1-v)\Lambda_{\mathsf P}^2 + (M_h^2-M_u^2)\left(x - \tfrac{1}{2}[1-w][1-v]\right) \\
&\qquad  + ( x [x-1] + \tfrac{1}{4} [1-v] [1-w^2]) \, m_{\mathsf P}^2\,;
\label{Omega}\\
\nonumber
{\mathpzc X}_{\mathsf P}(x;\sigma_\perp^2)  & =
\left[\int_{-1}^{1-2x} \! dw \int_{1+\frac{2x}{w-1}}^1 \!dv \right.\\
& \left. \quad + \int_{1-2x}^1 \! dw \int_{\frac{w-1+2x}{w+1}}^1 \!dv \right]
\frac{\rho_{\mathsf P}(w) }{{\mathpzc n}_{\mathsf P} } \frac{\Lambda_{\mathsf P}^2}{\sigma_\perp^2}\,; \label{X2c}
\end{align}
\end{subequations}
$\Lambda_{\mathsf P}$ is a mass-dimension parameter; and ${\mathpzc n}_{\mathsf P}$ is the canonical normalisation constant for the Bethe-Salpeter wave function \cite{Nakanishi:1969ph}.

The keystone in the bridge from Eq.\,\eqref{psiLFWF} to a realistic LFWF is the spectral weight, $\rho_H(w)$.  It was shown in Ref.\,\cite{Xu:2018eii} that when this is chosen judiciously, then results can be obtained for numerous hadron structure measures that are pointwise equivalent to the most advanced predictions currently available.  For pions and kaons, an optimal parametrisation is \cite{Xu:2018eii}:
\begin{align}
\rho_{\mathsf P}(\omega) 
& = \frac{1+\omega\; v_{\mathsf P}}{2a_{\mathsf P} b_0^{\mathsf P}} \left[\mbox{sech}^2 \left(\frac{\omega-\omega_0^{\mathsf P}}{2b_0^{\mathsf P}}\right)
 +\mbox{sech}^2 \left(\frac{\omega+\omega_0^{\mathsf P}}{2b_0^{\mathsf P}}\right)\right]
\,,
\label{eq:spectralw}
\end{align}
where $b_0^{\mathsf P}$, $\omega_0^{\mathsf P}$ control the density's profile, $v_{\mathsf P}$ introduces skewing driven by differences in the current-masses of a meson's valence-quark/-antiquark constituents, and $a_{\mathsf P}$ is a derived constant that fixes unit normalisation of the density.  This is a compact, yet flexible form; and using the parameters in Table~\ref{tab:params}, one reproduces the $\pi$ and $K$ DAs described in Refs.\,\cite{Cui:2020dlm, Cui:2020tdf}.

\smallskip

\noindent\textbf{4.$\;$Factorised \emph{Ans\"atze} for Light-Front Wave Functions}.
%
A simpler approach to the representation of $\pi$ and $K$ LFWFs is to employ a factorised (separable) \emph{Ansatz}:
\begin{equation}
\label{eq:facLFWF}
\psi_{{\mathsf P}}^{u}\left(x,k_\perp^2;\zeta_{\mathpzc H} \right)   =  \tilde\varphi_{\mathsf P}^u(x;\zeta_{\mathpzc H})  \tilde\psi_{{\mathsf P}}^{u}\left(k_\perp^2;\zeta_{\mathpzc H} \right) \,,
\end{equation}
where $\tilde\varphi_{\mathsf P}^u(x;\zeta_{\mathpzc H}) = \varphi_{\mathsf P}^u(x;\zeta_{\mathpzc H})/{\mathpzc n}_{\tilde\varphi_{\mathsf P}^u}$,
${\mathpzc n}_{\tilde\varphi_{\mathsf P}^u}^2 = \int_0^1 dx\, [\varphi_{\mathsf P}^u(x;\zeta_{\mathpzc H})]^2$,
%
%
and the optimal choice for $\tilde\psi_{{\mathsf P}}^{u}\left(k_\perp^2;\zeta_{\mathpzc H} \right)$ is influenced by the application. This exploits analyses in Ref.\,\cite{Xu:2018eii}, further expounded in Ref.\,\cite{Roberts:2021ppnp}, which reveal that when the LFWF has fairly uniform support, Eq.\,\eqref{eq:facLFWF} supplies an approximation, for use in calculating meson structure measures, whose accuracy exceeds the precision of foreseeable experiments.

Using Eq.\,\eqref{eq:uMzetaH}, Eq.\,\eqref{eq:facLFWF} entails
\begin{equation}
\label{PDFeqPDA2}
{\mathpzc u}^{\mathsf P}(x;\zeta_{\mathpzc H})
= [\tilde\varphi_{{\mathsf P}}^u(x;\zeta_{\mathpzc H})]^2 .
\end{equation}
Hence, the $x$-dependence of the factorised LFWFs is completely determined by sound forms for the $\pi$ and $K$ DFs at the hadron-scale.  Such are determined in Refs.\,\cite{Cui:2020dlm, Cui:2020tdf}:
\begin{align}\label{eq:param}
{\mathpzc u}^{\mathsf P}(x;\zeta_{\mathpzc H}) & = n_{\mathsf P} x^2 (1-x)^2 \nonumber \\
& \times \left[ 1 + \rho_{\mathsf P} x^\frac{\alpha_{\mathsf P}}{2} (1-x)^\frac{\beta_{\mathsf P}}{2} + \gamma_{\mathsf P} x^{\alpha_{\mathsf P}} (1-x)^{\beta_{\mathsf P}} \right]^2 \,,
\end{align}
completed by the entries in Table~\ref{tab:DFs}.
Only the $k_\perp^2$-dependence remains to be determined.  As shown next, that can be done by using one or two extra pieces of empirical information.

\begin{table}[t!]
\begin{center}
\begin{tabular}{l|c|c|c|c|c}\hline
${\mathsf P}$ & $ n_{\mathsf P}$ & $\rho_{\mathsf P}$ & $\gamma_{\mathsf P}$ & $\alpha_{\mathsf P}$ & $\beta_{\mathsf P}$ \\ \hline
$\pi$ & $375.3$ & $-2.51$ & $\phantom{-}2.03$ & $1.0\phantom{00}$ & $1.0\phantom{00}$ \\\hline
$K$ &  $299.2$ & $\phantom{-}5.00$ & $-5.97$ & $0.064$ & $0.048$ \\\hline
\end{tabular}
\end{center}
\caption{
\label{tab:DFs}
When these powers and coefficients are used in Eq.\,\eqref{eq:param}, one obtains representations of $\pi$ and $K$ DFs that are in accord with all available data \cite{Cui:2020dlm, Cui:2020tdf}.
}
\end{table}

\smallskip

\noindent\textbf{5.$\;$Overlap Generalised Parton Distribution}.
Working with the LFWFs just described, one can extend the analysis of one-dimensional measures of $\pi$ and $K$ structure in Refs.\,\cite{Cui:2020dlm, Cui:2020tdf} and develop three-dimensional images by employing the following overlap representation of GPDs \cite{Burkardt:2002hr, Diehl:2003ny}:
%
\begin{equation}
%
H^u_{\mathsf P}(x,\xi,t;\zeta_{\mathpzc H})  =
 \int \frac{d^2{k_\perp}}{16 \pi^3}
\psi_{{\mathsf P}}^{u\ast}\left(x_-,{k}_{\perp -}^2;\zeta_{\mathpzc H} \right)
\psi_{{\mathsf P}}^{u}\left( x_+,{k}_{\perp +}^2;\zeta_{\mathpzc H} \right) \,,
\label{eq:overlap}
\end{equation}
%
where: $P=(p^\prime+p)/2$, with $p^\prime$, $p$ being the final, initial meson momenta in the defining scattering process; $\Delta = p^\prime - p$, $P\cdot \Delta = 0$, $t=-\Delta^2$; the ``skewness'' $\xi = [-n\cdot \Delta]/[2 n\cdot P]$, \mbox{$|\xi|\leq 1$}; and
$x_\mp = (x \mp \xi)/(1\mp \xi)$,
${k}_{\perp \mp} = {k}_\perp \pm ({\Delta}_\perp/2)(1-x)/(1 \mp \xi) $\,.
%
%
%
Symmetries of Nature guarantee $H^u_{\mathsf P}(x,\xi,t)  =H^u_{\mathsf P}(x,-\xi,t)$; hence, in the following, we only consider $\xi \geq 0$.

Eq.\,\eqref{eq:overlap} defines the GPD on $|x|\geq \xi$, referred to as the DGLAP domain because scale evolution thereupon is implemented via the equations described in Refs.\,\cite{Dokshitzer:1977, Gribov:1972, Lipatov:1974qm, Altarelli:1977}.  Additionally, given that the quark GPD is only nonzero on $x \geq -\xi $, with the antiquark GPD nonzero on $x\leq\xi$, Eq.\,\eqref{eq:uMzetaH} is seen to follow as the GPD's forward ($\Delta^2=0$) limit.  This is a general result.

A complete definition of the GPD requires extension of Eq.\,\eqref{eq:overlap} onto $|x| \leq \xi$, whereupon the initial and final state LFWFs differ by two in their parton content.  This is known as the ERBL domain because scale evolution is implemented via the equations described in Refs.\,\cite{Lepage:1979zb, Efremov:1979qk, Lepage:1980fj}.   In attempting such an extension, many challenges are encountered, but progress is being made \cite{Chouika:2017dhe, Chouika:2017rzs}.


%
%
%
\smallskip

\noindent\textbf{6.$\;$Potential for Insights from Factorised Wave Functions}.
Inserting Eq.\,\eqref{eq:facLFWF} into Eq.\,\eqref{eq:overlap} and using Eq.\,\eqref{PDFeqPDA2}, one obtains
\begin{subequations}
\label{eq:HfacT}
\begin{align}
H^u_{\mathsf P}(x,&\xi,-\Delta_\perp^2;\zeta_{\mathpzc H})  \nonumber\\
& = \Theta(x_-) \sqrt{{\mathpzc u}^{\mathsf P}\left(x_-;\zeta_{\mathpzc H} \right) {\mathpzc u}^{\mathsf P}\left(x_+;\zeta_{\mathpzc H} \right)} \; \Phi_{\mathsf P}^u\left( z; \zeta_{\mathpzc H} \right) \,,
\label{eq:Hfac} \\
%
\Phi_{\mathsf P}^u (z;&\zeta_{\mathpzc H}) =  \int \frac{d^2{k_\perp}}{16 \pi^3}
 \tilde \psi_{{\mathsf P}}^{u}\left({k}_\perp^2;\zeta_{\mathpzc H} \right)
 \tilde \psi_{{\mathsf P}}^{u}\left(\left({k}_\perp - {s}_\perp \right)^2;\zeta_{\mathpzc H} \right) \,,
 \label{eq:Phi}
\end{align}
\end{subequations}
where $\Theta$ is the Heaviside function,
$z={s}_\perp^2={\Delta}_\perp^2 (1-x)^2 /(1-\xi^2)^2$,
and canonical normalisation 
guarantees $\Phi_{\mathsf P}^u(0;\zeta_{\mathpzc H})=1$.

One of the many implicit GPD features is the fact that a given quark's contribution to the target meson's elastic electromagnetic form factor is obtained by calculating a zeroth moment:
\begin{equation}\label{eq:FF}
F_{\mathsf P}^u(\Delta^2) = \int_{-1}^{1} dx \, H_{\mathsf P}^u(x,0,-\Delta^2;\zeta_{\mathpzc H})  \,.
\end{equation}

Differentiating the GPD specified by Eqs.\,\eqref{eq:HfacT}, one obtains
\begin{subequations}
\label{eq:PhifromFA}
\begin{align}
\label{eq:PhifromF}
\frac{\partial^n}{\partial^n z} \left. \Phi_{\mathsf P}^u(z;\zeta_{\mathpzc H}) \right|_{z=0} & =
\frac{1}{\langle x^{2n}\rangle_{\bar{h}}^{\zeta_{\mathpzc H}} }
\left. \frac{d^n F_{\mathsf P}^u(\Delta^2)}{d(\Delta^2)^n}  \right|_{\Delta^2=0} \,, \\
\langle x^{2n} \rangle_{\bar{h}}^{\zeta_{\mathpzc H}}  = \langle (1-x)^{2n} \rangle_{u}^{\zeta_{\mathpzc H}}
& = \int_0^1 dx (1-x)^{2n} {\mathpzc u}^{\mathsf P}(x;\zeta_{\mathpzc H}) \,.
\end{align}
\end{subequations}
Plainly, when using a factorised \emph{Ansatz} for the LFWF, the $k_\perp^2$-overlap portion of the $u$-in-${\mathsf P}$ GPD is completely determined by the $u$-quark's DF and its contribution to the elastic form factor.

Now consider the meson's complete elastic form factor:
\begin{equation}\label{eq:FFM}
F_{\mathsf P}(\Delta^2) = e_u F_{\mathsf P}^u(\Delta^2) + e_{\bar h} F_{\mathsf P}^h(\Delta^2) \,,
\end{equation}
with $e_u$, $e_{\bar h}$ the valence-constituent electric charges in units of the positron charge.  The meson's charge radius is defined via $r^2_{\mathsf P} = -\left.[6/F_{\mathsf P}(0)]\, dF_{\mathsf P}(\Delta^2)/d\Delta^2 \right|_{\Delta^2=0}$.  Hence, using Eqs.\,\eqref{eq:PhifromFA}:
\begin{subequations}
\label{eq:dphiMdz}
\begin{align}
& \left. \frac{\partial}{\partial z} \Phi_{\mathsf P}^u(z;\zeta_{\mathpzc H}) \right|_{z=0} = -\frac{r^2_{\mathsf P}}
{4{\mathpzc x}_{\mathsf P}^2(\zeta_{\mathpzc H})} \,,
\label{eq:dphiMKdzu} \\
& \left.\frac{\partial}{\partial z} \Phi_{\mathsf P}^{\bar h}(z;\zeta_{\mathpzc H}) \right|_{z=0} =  \left. (1 - {\mathpzc d}_{\mathsf P} ) \frac{\partial}{\partial z} \Phi_{\mathsf P}^u(z;\zeta_{\mathpzc H}) \right|_{z=0}   \,, \label{eq:dphiMdzh}
\end{align}
\end{subequations}
where
\begin{equation}
\label{definex}
{\mathpzc x}_{\mathsf P}^2(\zeta_{\mathpzc H}) = \langle x^2 \rangle_{\bar{h}}^{\zeta_{\mathpzc H}} + \tfrac{1}{2} (1 - {\mathpzc d}_{\mathsf P}) \langle x^2 \rangle_u^{\zeta_{\mathpzc H}}\,,
\end{equation}
and ${\mathpzc d}_{\mathsf P} \propto (M_{\bar h}-M_u )$ is introduced to express the impact of any mass-difference between the valence constituents on the meson's charge distribution.  ${\mathpzc d}_\pi=0$ in the limit of isospin symmetry; and for the kaon, $M_{\bar h}>M_u$ and  ${\mathpzc d}_{K}>0$ because the $\bar s$-quark contribution to $F_K$ is harder than that of the $u$-quark \cite{Chen:2012txa, Gao:2017mmp}.

Additional insights may be gained by using a Gaussian to describe the $k_\perp^2$-dependence of the LFWF in Eq.\,\eqref{eq:facLFWF}.  Then Eqs.\,\eqref{eq:dphiMdz} completely constrain the pointwise behaviour:
\begin{equation}\label{eq:LFWFgauss}
\psi_{{\mathsf P}}^{u}\left(x,k_\perp^2;\zeta_{\mathpzc H} \right)  = \left(\frac{16 \pi^2 r_{\mathsf P}^2}{{\mathpzc x}_{\mathsf P}^2(\zeta_{\mathpzc H})} \, u^{\mathsf P}(x;\zeta_{\mathpzc H})\right)^{1/2}
\exp{\left(-\frac{ r_{\mathsf P}^2 k_\perp^2}{2 {\mathpzc x}_{\mathsf P}^2(\zeta_{\mathpzc H})}\right)}  \,;
\end{equation}
and the DGLAP-domain GPD is
\begin{align}
H_{\mathsf P}^u(x,\xi,-\Delta^2;\zeta_{\mathpzc H}) &=  \Theta(x-\xi) \sqrt{{\mathpzc u}^{\mathsf P}\left(x_-;\zeta_{\mathpzc H} \right) {\mathpzc u}^{\mathsf P}\left(x_+;\zeta_{\mathpzc H} \right)}
\nonumber \\
& \qquad \times
\exp{\left( - \frac{\Delta^2 \, r_{\mathsf P}^2 (1-x)^2}{4 {\mathpzc x}_{\mathsf P}^2(\zeta_{\mathpzc H})(1-\xi^2) }\right)} \,.
\label{eq:gaussH}
\end{align}
The properties of meson bound-states under charge conjugation entail that the $\bar{h}$-in-${\mathsf P}$ GPD is obtained by replacing $\Theta(x-\xi) \to -\Theta(-x-\xi)$, ${\mathpzc u}^{\mathsf P} \to \bar{{\mathpzc s}}^{\mathsf P}$, $r_{\mathsf P}^2 \to r_{\mathsf P}^2 (1-{\mathpzc d}_H)$ and $x \to |x|$ in Eq.\,\eqref{eq:gaussH}.

Exploiting these features, one can compute the charged-kaon form factor using Eq.\,\eqref{eq:FFM} with the PTIR kaon LFWF specified in Sec.\,3.  The charge-radius calculated from that form factor is $r_K^2 = (0.56\,{\rm fm})^2$.  (Experiment \cite{Zyla:2020zbs}: $(0.560(31)\,{\rm fm})^2$.)  The factorised \emph{Ansatz} reproduces this computed radius with ${\mathpzc d}_K=0.07$ in Eqs.\,\eqref{eq:dphiMdz}, \eqref{eq:LFWFgauss}.

\begin{figure}[t]
\vspace*{3.5ex}

\leftline{\hspace*{0.5em}{\large{\textsf{A}}}}
\vspace*{-5ex}
\centerline{\includegraphics[clip, width=0.36\textwidth]{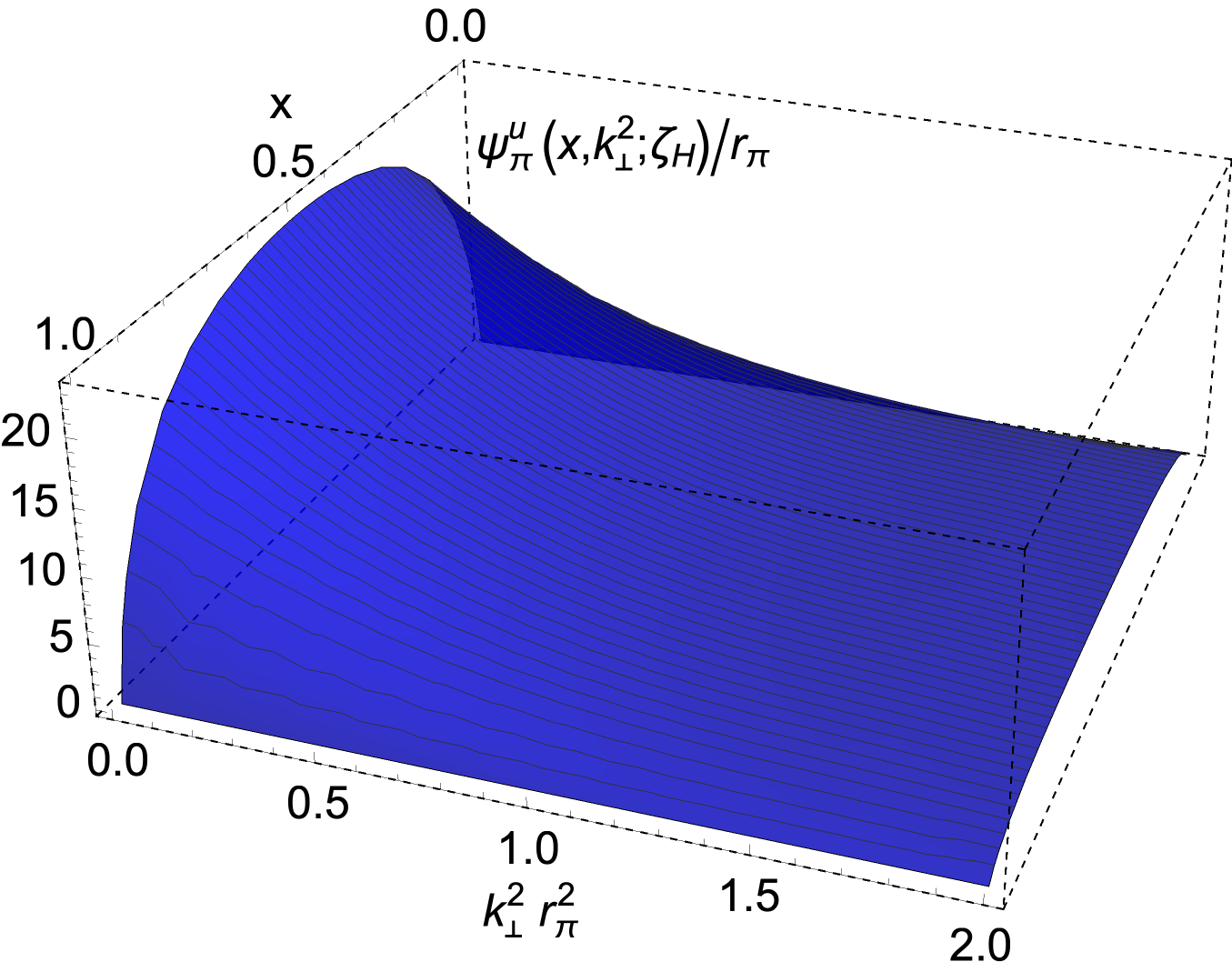}}
\vspace*{6ex}

\leftline{\hspace*{0.5em}{\large{\textsf{B}}}}
\vspace*{-5ex}
\centerline{\includegraphics[clip, width=0.37\textwidth]{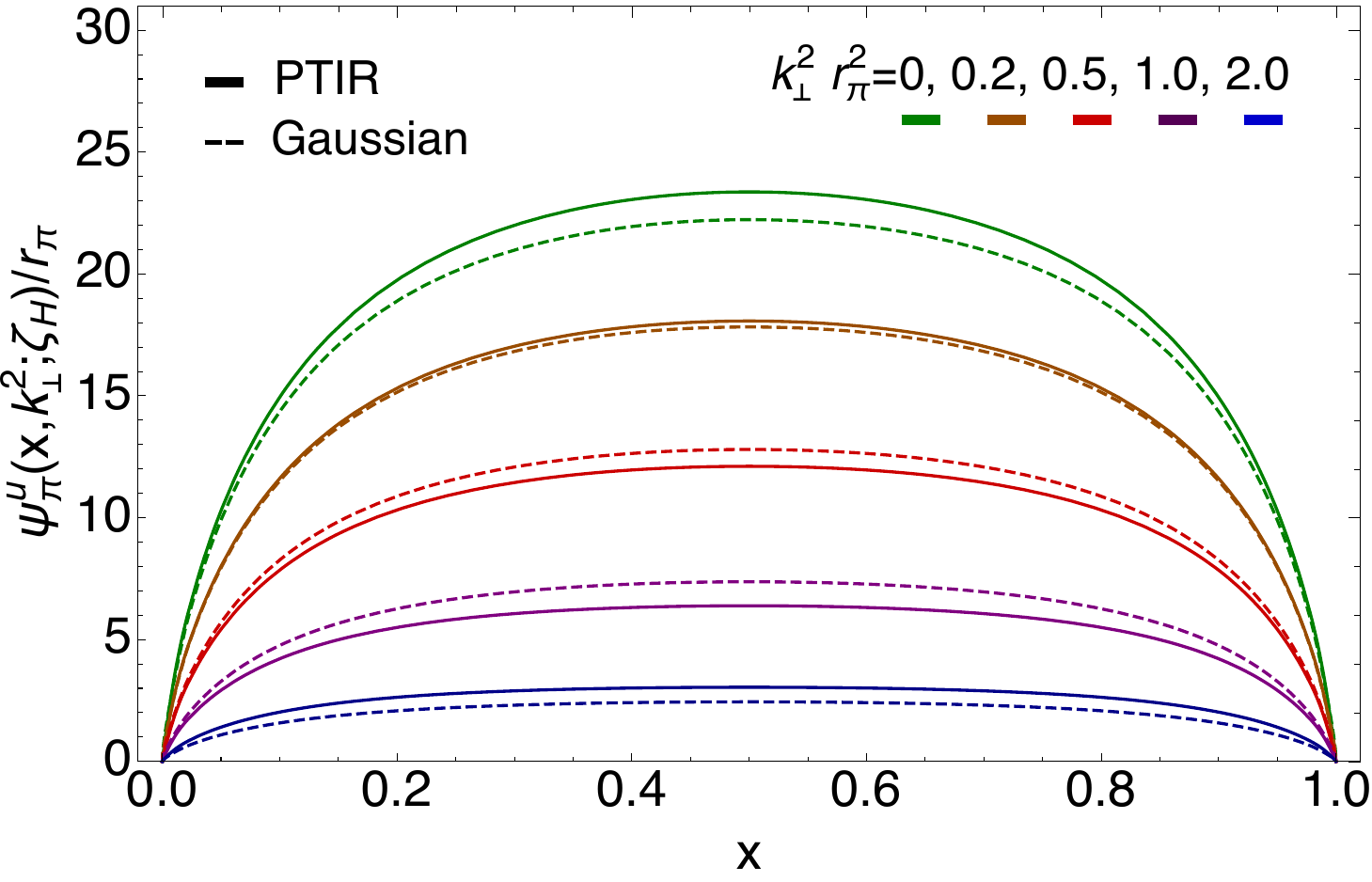}}
%
\caption{
\emph{Upper panel}\,--\,{\sf A}.
Pion PTIR LFWF defined by Eqs.\,\eqref{psiLFWF} -- \eqref{eq:spectralw} and Table~\ref{tab:params}, using which $r_\pi \approx 0.69\,$fm.  (Experiment \cite{Zyla:2020zbs}: $0.659(4)\,$fm.)
\emph{Lower panel}\,--\,{\sf B}.
LFWF in Panel\,A (solid curves) compared with the factorised \emph{Ansatz}, Eq.\,\eqref{eq:LFWFgauss} (dashed curves), drawn as a function of $x$ on slices of constant $k_\perp^2 r_\pi^2$.
\label{fig:piLFWF}}
\end{figure}


\begin{figure}[t]
\vspace*{3.5ex}

\leftline{\hspace*{0.5em}{\large{\textsf{A}}}}
\vspace*{-5ex}
\centerline{\includegraphics[clip, width=0.37\textwidth]{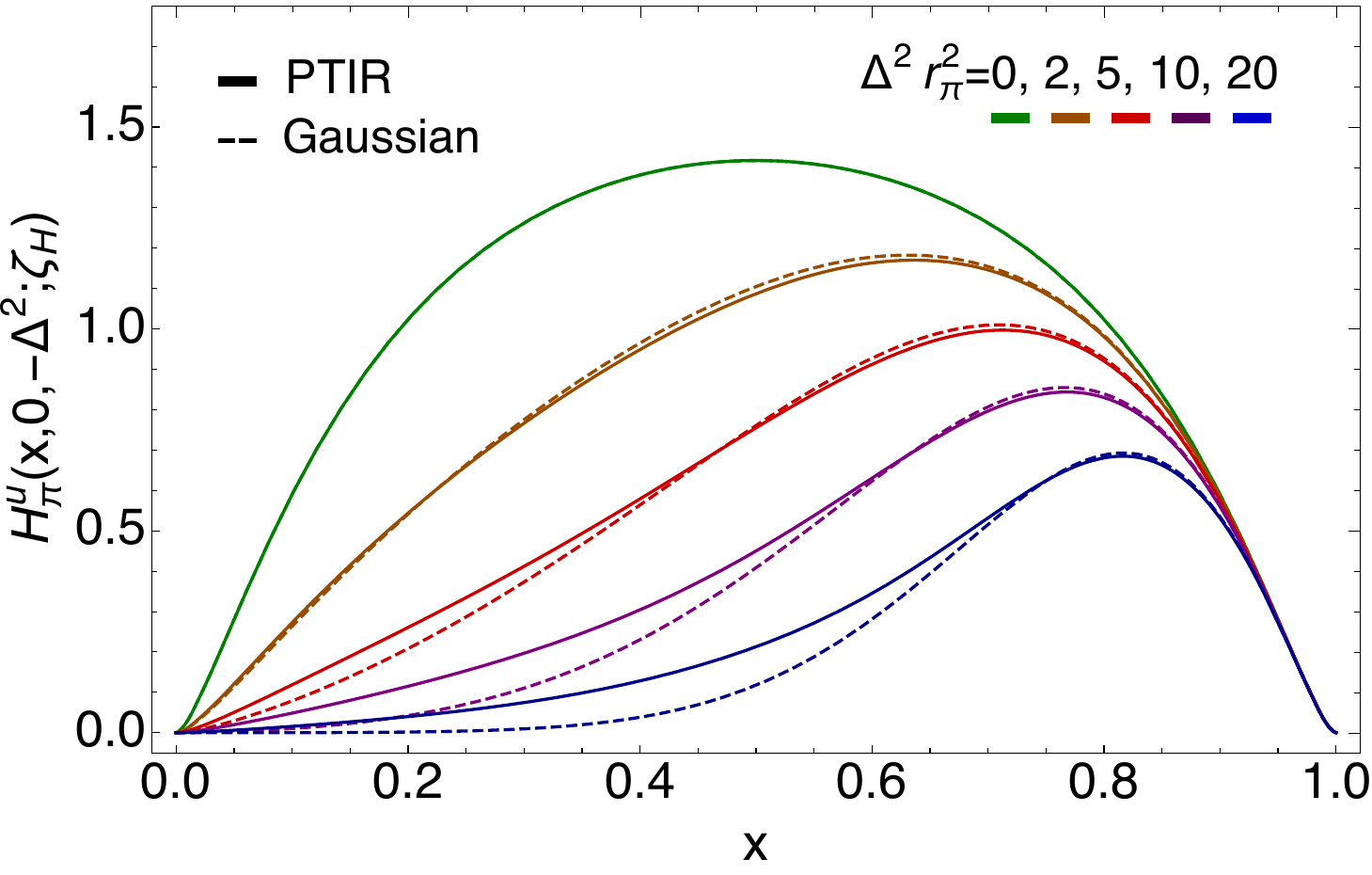}}
\vspace*{6ex}

\leftline{\hspace*{0.5em}{\large{\textsf{B}}}}
\vspace*{-5ex}
\centerline{\includegraphics[clip, width=0.37\textwidth]{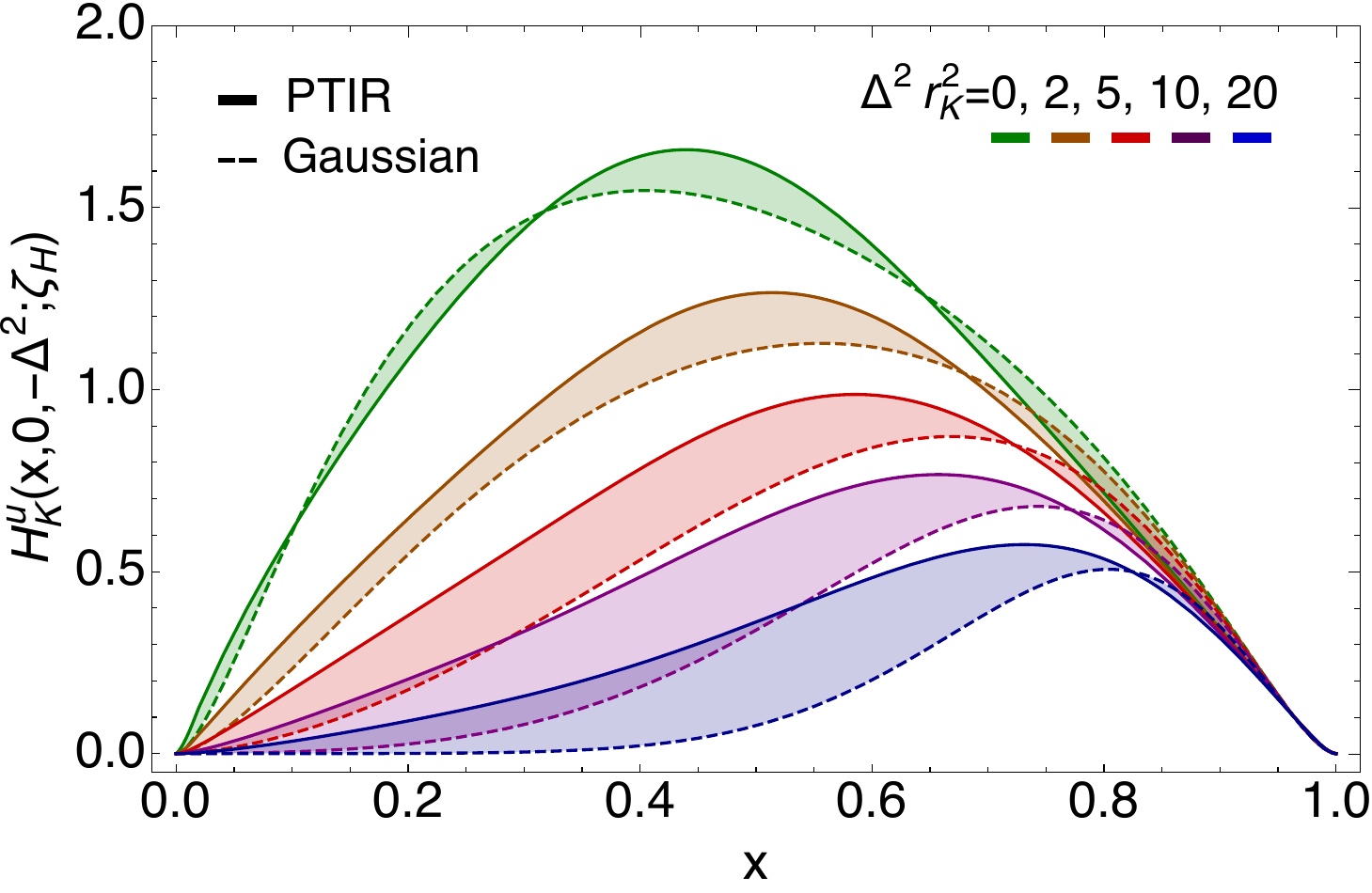}}
\caption{
\label{fig:GPDs}
\emph{Upper panel}\,--\,{\sf A}.  $u$-in-$\pi$ GPD, drawn as a function of $x$ on slices of constant $\Delta^2 r_\pi^2$: solid curves -- obtained using PTIR LFWF, Eqs.\,\eqref{psiLFWF} -- \eqref{eq:spectralw} and Table~\ref{tab:params}; and dashed curves -- results using factorised LFWF, Eq.\,\eqref{eq:gaussH}.
\emph{Lower panel}\,--\,{\sf B}.  Analogous curves for $u$-in-$K$ GPD.  (Shading highlights the curves that should be compared.)   The factorised $K$ LFWF is obtained with ${\mathpzc d}_K = 0.07$ in Eqs.\,\eqref{eq:dphiMdz}, \eqref{eq:LFWFgauss}.
}

\end{figure}

The PTIR pion LFWF defined by Eqs.\,\eqref{psiLFWF} -- \eqref{eq:spectralw} and Table~\ref{tab:params} is depicted in Fig.\,\ref{fig:piLFWF}A.  It is compared with the factorised LFWF, Eq.\,\eqref{eq:gaussH}, in Fig.\,\ref{fig:piLFWF}B.  Plainly, for practical purposes, the factorised \emph{Ansatz} provides a pointwise satisfactory approximation; hence, can yield sound insights.

The comparisons are extended to the associated $\pi$ and $K$ GPDs in Fig.\,\ref{fig:GPDs}.
Fig.\,\ref{fig:GPDs}A reveals that the factorised \emph{Ansatz} supplies a good approximation for the $\pi$ on the entire depicted domain of $\Delta^2$, although the pointwise accuracy degrades slowly with increasing $\Delta^2$.
Turning to the $K$, examination of Eqs.\,\eqref{psiLFWF}, \eqref{X2c}, as in Ref.\,\cite{Xu:2018eii}, reveals that pointwise discrepancies between a LFWF and a well-developed factorised approximation can grow with $m_{\mathsf P}^2$, $M_{\bar h}^2-M_u^2$.  Hence, regarding Fig.\,\ref{fig:GPDs}B, the pointwise agreement between the PTIR GPD and that produced by the factorised LFWF, Eq.\,\eqref{eq:LFWFgauss}, does not match that for the pion.  Nevertheless, the semiquantitative accord indicates that even the simple factorised \emph{Ansatz} can yield a fair picture.

It is worth stressing two qualitative aspects of the hadron-scale $\pi$ and $K$ GPDs in Fig.\,\ref{fig:GPDs}.
Recalling that $H_{\mathsf P}^u(x,0,0;\zeta_{\mathpzc H}) = {\mathpzc u}^{\mathsf P}(x;\zeta_{\mathpzc H})$, one is reminded that the maximum of $H_\pi^u(x,0,0;\zeta_{\mathpzc H})$ lies at $x=1/2$ and that of $H_K^u(x,0,0;\zeta_{\mathpzc H})$ at $x=0.4$ \cite{Cui:2020dlm}.  This 20\% shift is typical for Higgs-boson modulation of EHM.
The peak position changes as the momentum transfer to the target increases: for both the $\pi$ and $K$, as $\Delta^2$ grows, the GPD's support shifts toward $x=1$, becoming ever more localised in the neighbourhood of its peak value; hence, focused more tightly within the valence domain.
This latter remark emphasises that it is valence partons which are revealed by hard probes.

\begin{figure}[t]
\vspace*{3.5ex}

\leftline{\hspace*{0.5em}{\large{\textsf{A}}}}
\vspace*{-5ex}
\centerline{\includegraphics[clip, width=0.36\textwidth]{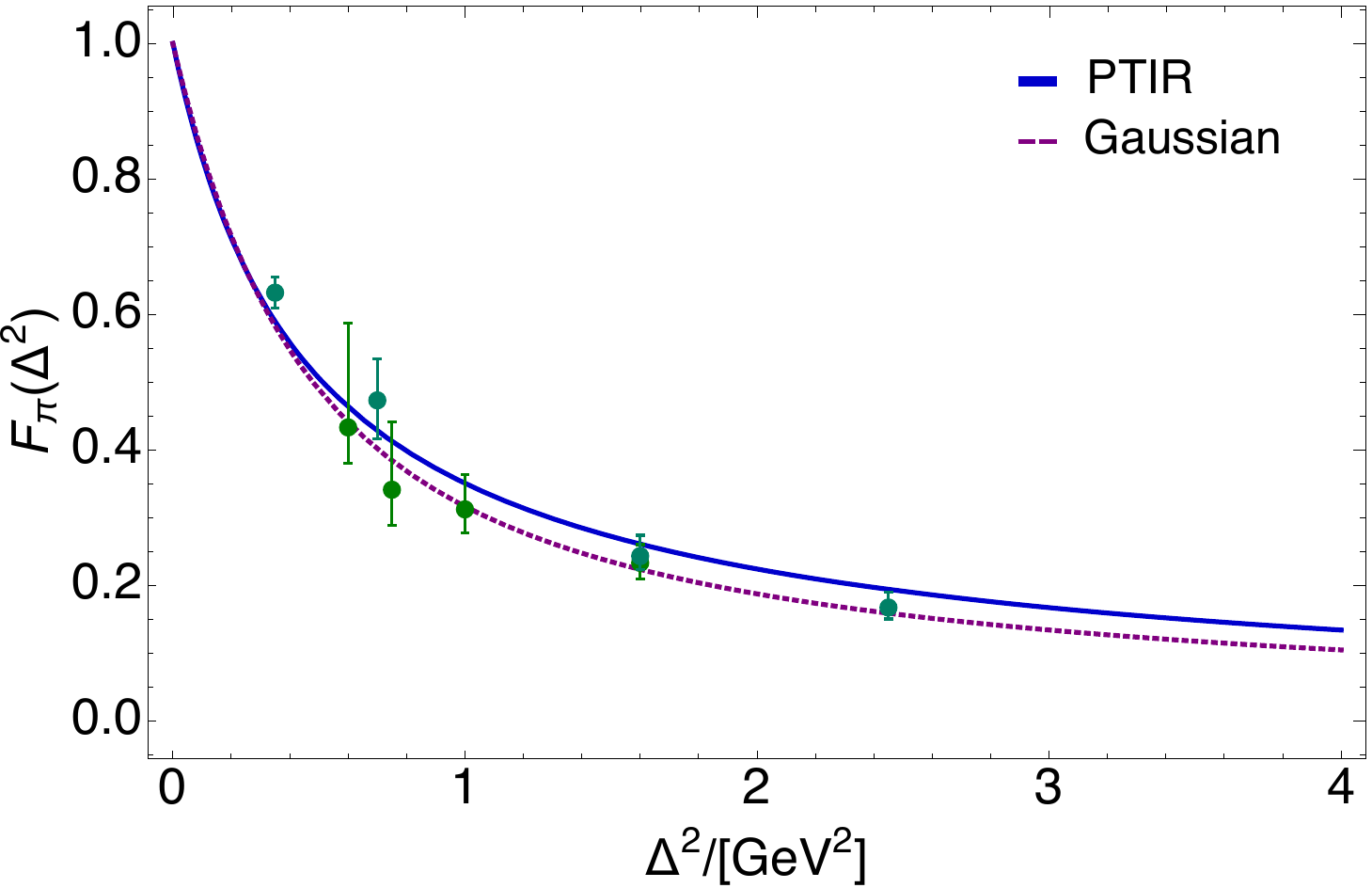}}
\vspace*{6ex}

\leftline{\hspace*{0.5em}{\large{\textsf{B}}}}
\vspace*{-5ex}
\centerline{\includegraphics[clip, width=0.36\textwidth]{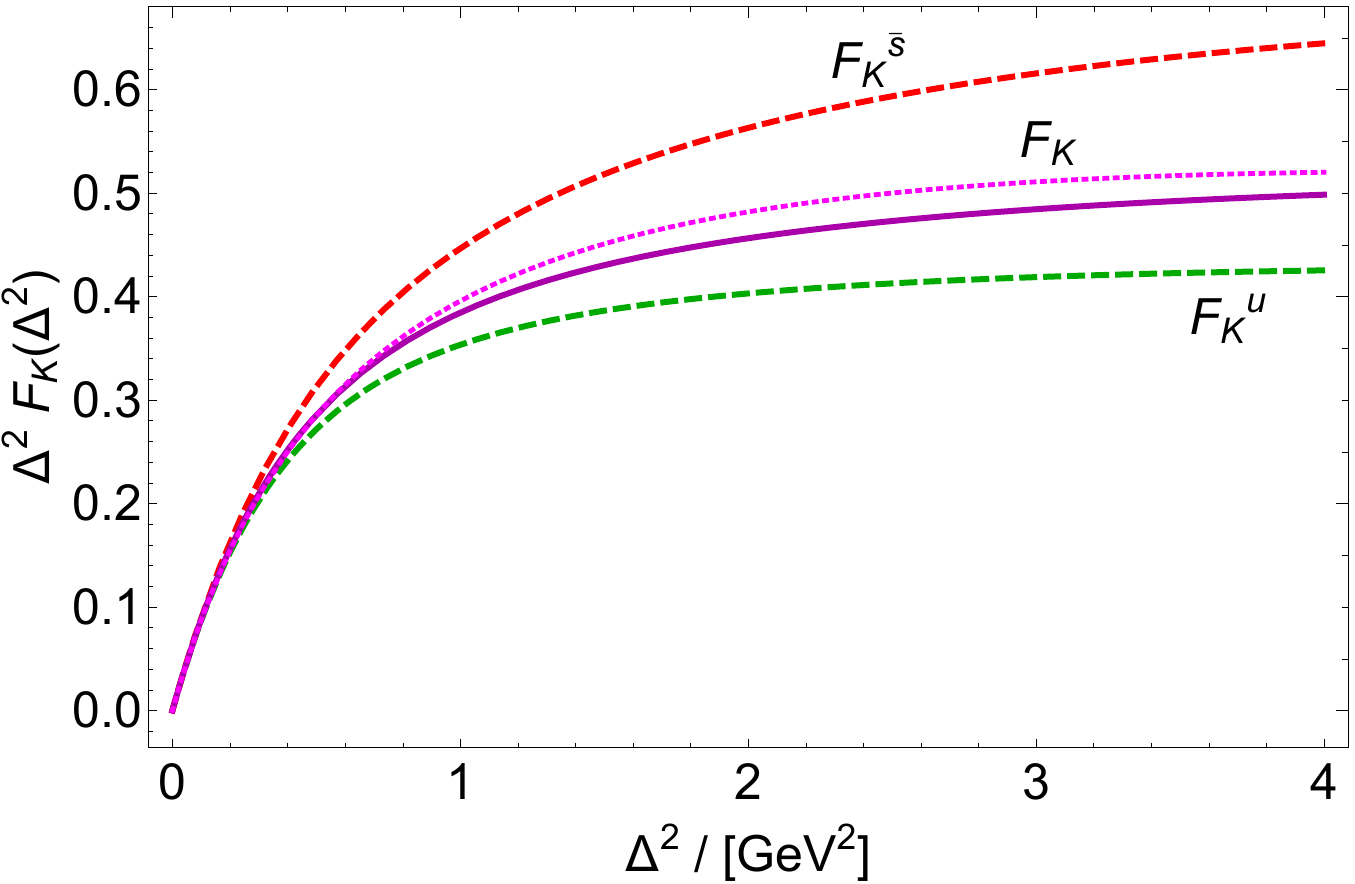}}
%
\caption{
\emph{Upper panel}\,--\,{\sf A}.  $F_\pi(\Delta^2)$ obtained using the GPDs compared in Fig.\,\ref{fig:GPDs}: PTIR -- solid blue curve; and factorised LFWF -- short-dashed magenta.  Data from Ref.\,\cite{Huber:2008id}.
\emph{Lower panel}\,--\,{\sf B}.  $\Delta^2 F_K(\Delta^2)$ obtained using PTIR $K$ GPD, including a comparison with the prediction in Ref.\,\cite{Gao:2017mmp} -- dotted magenta curve.  (Flavour-separated form factor contributions normalised to unity at $\Delta^2=0$.)
\label{fig:FFs}}
\end{figure}

Pion and kaon elastic electromagnetic form factors, computed using Eq.\,\eqref{eq:FF} and its analogue for the $K^+$, are drawn in Fig.\,\ref{fig:FFs}.
Regarding Fig.\,\ref{fig:FFs}A, it is clear that the PTIR and factorised-\emph{Ansatz} GPDs are practically indistinguishable for the $\pi$ and both agree with existing data \cite{Huber:2008id}.  (\emph{N.B}.\ These data were not used to constrain the PTIR LFWFs, which are completely fixed by the meson DAs determined in Refs.\,\cite{Cui:2020dlm, Cui:2020tdf}.)
Fig.\,\ref{fig:FFs}B depicts $\Delta^2 F_{K^+}(\Delta^2)$ and its flavour separation as computed using the PTIR GPD.  This simple PTIR GPD delivers agreement with the charged-kaon form factor obtained in Ref.\,\cite{Gao:2017mmp}, which employed a far more elaborate and computationally intensive approach.  Similarly, there is agreement between our results for the independent contributions from the $u$- and $\bar s$-quarks and those predicted in Ref.\,\cite{Gao:2017mmp}, \emph{e.g}.\ at $\Delta^2=4\,{\rm GeV}^2$, $F_{K^+}^{\bar s}/F_{K^+}^{u}  = 1.5$.  As $\Delta^2\to \infty$, this ratio approaches unity.


\smallskip

\noindent\textbf{7.$\;$GPDs in Impact Parameter Space}.
Arranging kinematics such that $\Delta^2>0$, $\Delta\cdot P_{\mathsf P} = 0$, $\xi=0$, and calculating a two-dimensional Fourier transform of the GPD with respect to the remaining two degrees-of-freedom, one obtains an impact parameter space GPD:
\begin{equation}
{\mathpzc u}^{\mathsf P}(x,b_\perp^2;\zeta_{\mathpzc H}) = \int_0^\infty \frac{d\Delta}{2\pi} \Delta J_0(b_\perp \Delta) \, H_{\mathsf P}^u(x,0,-\Delta^2;\zeta_{\mathpzc H})\,,
\label{eq:IPDHgen}
\end{equation}
where $J_0$ is a cylindrical Bessel function.  This density describes the probability of finding a parton within the light-front at a transverse position $b_\perp$ from the meson's centre of transverse momentum.

Introducing the general form of the factorised-\emph{Ansatz} GPD, Eq.\,\eqref{eq:HfacT}, into Eq.\,\eqref{eq:IPDHgen}, one obtains
\begin{equation}
{\mathpzc u}^{\mathsf P}(x, b_\perp^2;\zeta_{\mathpzc H}) =  \frac{{\mathpzc u}^{\mathsf P}(x;\zeta_{\mathpzc H}) }{(1-x)^2}  \int_0^\infty \frac{s ds}{2\pi} \, \Phi_{\mathsf P}(s^2;\zeta_{\mathpzc H}) J_0\left( \frac{b_\perp s}{1-x} \right) \,.
\label{eq:IPDH}
\end{equation}
Thus 
the light-front longitudinal distribution of the mean-square light-front transverse extent of $u$-in-${\mathsf P}$: 
%
%
\begin{equation}
\label{eq:b2x}
\langle b_\perp^2(x;\zeta_{\mathpzc H}) \rangle_{u}^{\mathsf P}
=r_{\mathsf P}^2  \frac{(1-x)^2 {\mathpzc u}^{\mathsf P}(x;\zeta_{\mathpzc H})}{{\mathpzc x}_{\mathsf P}^2(\zeta_{\mathpzc H})}  \,.
\end{equation}
%
Eq.\,\eqref{eq:b2x} makes an array of physical properties explicit, \emph{e.g}.\
using Eq.\,\eqref{eq:param} and Table~\ref{tab:DFs}, one finds that $\langle b_\perp^2(x;\zeta_{\mathpzc H})\rangle $ is broadest on the neighbourhood $x\simeq 0.23$ and becomes progressively narrower as $x\to 1$.  Under QCD evolution \cite{Dokshitzer:1977, Gribov:1972, Lipatov:1974qm, Altarelli:1977} to $\zeta>\zeta_{\mathpzc H}$, the peak in $\langle b_\perp^2(x;\zeta)\rangle$ moves to $x=0$ and increases in height, becoming unbounded, whilst support is simultaneously stripped from the valence-quark domain \cite{Raya:2021:progress}.
The ${\bar h}$ result is obtained by making the replacement ${\mathpzc u}^{\mathsf P}(x;\zeta_{\mathpzc H}) \to (1-{\mathpzc d}_{\mathsf P}) \bar{\mathpzc h}(x;\zeta_{\mathpzc H})$.

The mean-square light-front transverse extent is obtained by integrating Eq.\,\eqref{eq:b2x} over $x$, which yields:
\begin{subequations}
\label{EqmsqLFE}
\begin{align}
\langle b_\perp^2(\zeta_{\mathpzc H}) \rangle_{u}^\pi & = \frac{2}{3} r_\pi^2 = \langle b_\perp^2(\zeta_{\mathpzc H}) \rangle_{\bar d}^\pi \,,\\
\langle b_\perp^2(\zeta_{\mathpzc H}) \rangle_{u}^K & = 0.71 r_K^2\,,
\langle b_\perp^2(\zeta_{\mathpzc H}) \rangle_{\bar s}^K = 0.58 r_K^2\,.
\end{align}
\end{subequations}
Plainly, in a $u\bar h$ meson, there is a separation of baryon number, with the heavier $\bar h$-quark lying, on average, closer to the centre of transverse momentum.

\begin{figure}[t]
\vspace*{3.5ex}

\leftline{\hspace*{0.5em}{\large{\textsf{A}}}}
\vspace*{-5ex}
\centerline{\includegraphics[clip, width=0.4\textwidth]{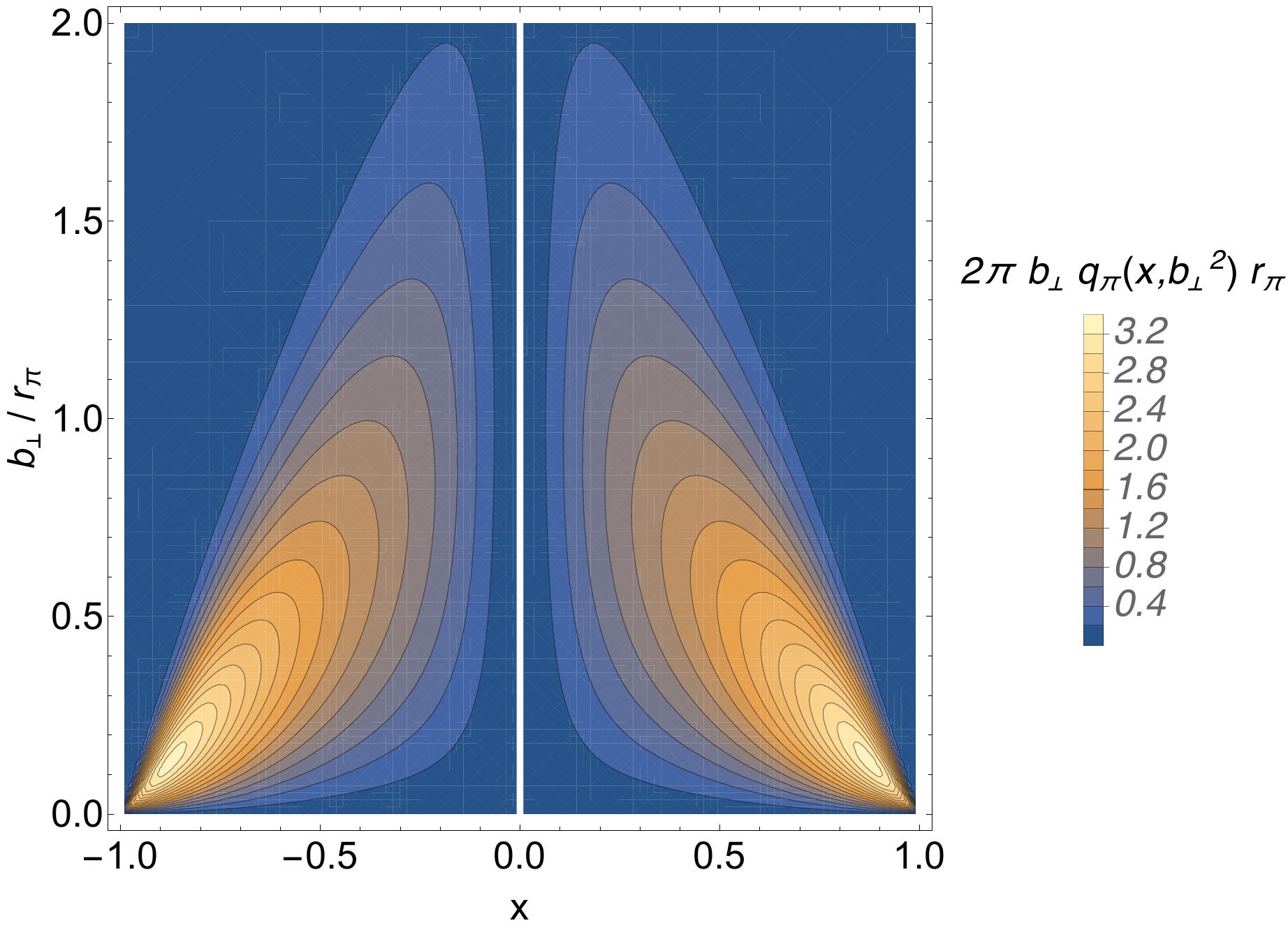}}
\vspace*{6ex}

\leftline{\hspace*{0.5em}{\large{\textsf{B}}}}
\vspace*{-5ex}
\centerline{\includegraphics[clip, width=0.4\textwidth]{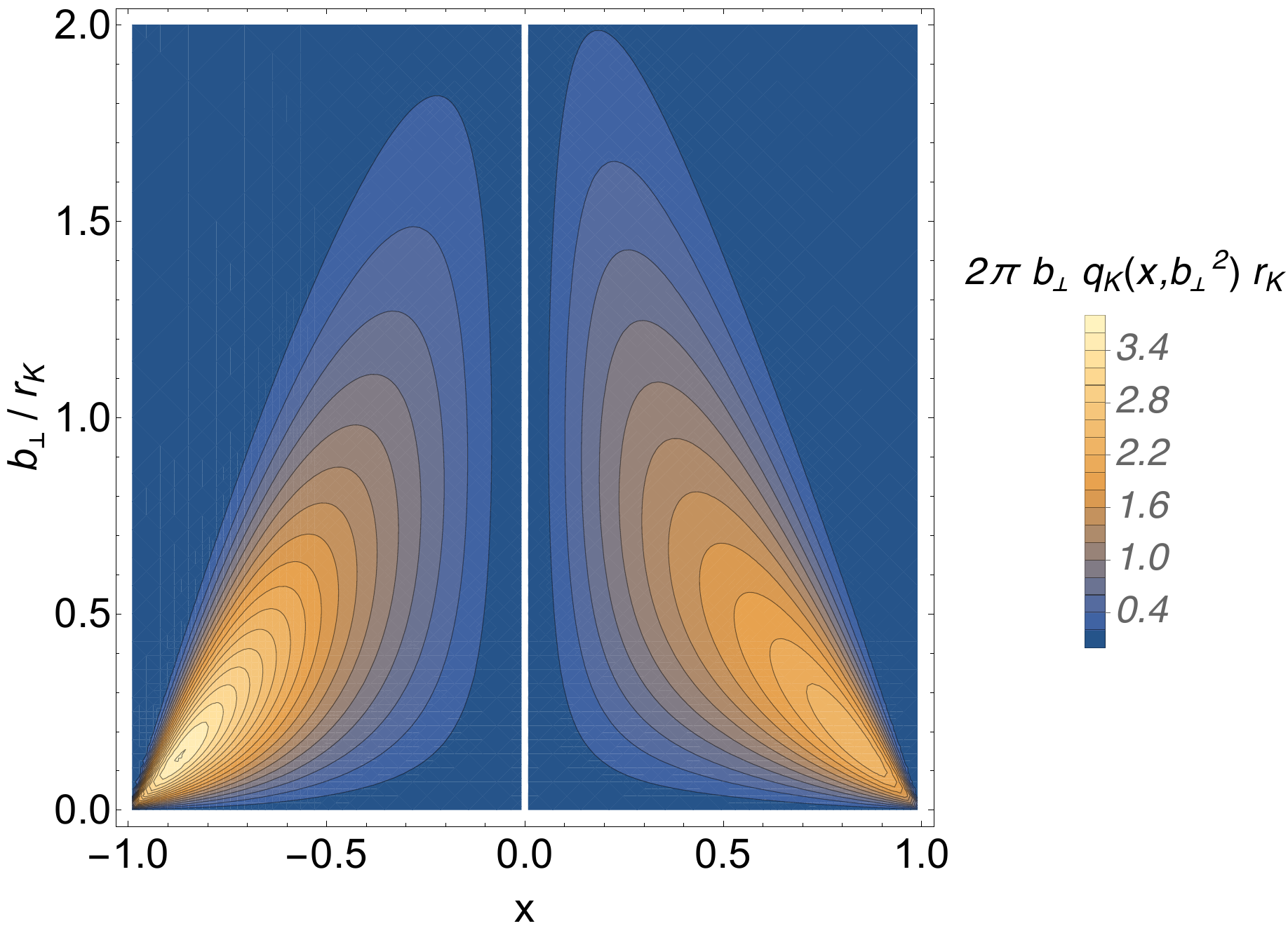}}
%
\caption{\label{fig:IPDGPDs} 
\emph{Upper panel}\,--\,{\sf A}.  Pion impact parameter space distributions for $\bar d$- (left) and $u$-quarks (right), expressing the $x$-dependent probability density associated with locating these objects at a distance $|b_\perp|$ from the pion's centre of transverse momentum.
\emph{Lower panel}\,--\,{\sf B}.  Similar profiles for $\bar s$ and $u$-quarks in the $K^+$.
Curves calculated using Eq.\,\eqref{eq:IPDHq} and its analogues (factorised \emph{Ans\"atze)}.
}

\end{figure}

Using the Gaussian \emph{Ansatz} defined by Eq.\,\eqref{eq:gaussH}, one obtains
\begin{equation}
{\mathpzc u}^{\mathsf P}(x,b_\perp^2;\zeta_{\mathpzc H}) =
\frac{{\mathpzc x}_{\mathsf P}^2(\zeta_{\mathpzc H})}{\pi r_{\mathsf P}^2} \, \frac{{\mathpzc u}^{\mathsf P}(x;\zeta_{\mathpzc H})}{(1-x)^2}
 \exp{\left(- \frac{{\mathpzc x}_{\mathsf P}^2(\zeta_{\mathpzc H})}{(1-x)^2} \frac{b_\perp^2}{r_{\mathsf P}^2} \right)} \,.
\label{eq:IPDHq}
\end{equation}
The ${\bar h}$-in-${\mathsf P}$ GPD, which has nonzero support on $x\in (-1,0)$, is obtained by making the following replacements in Eq.\,\eqref{eq:IPDHq}: ${\mathpzc u}^{\mathsf P} \to \bar{{\mathpzc h}}^{\mathsf P}$, $x \to |x|$ and $r_{\mathsf P}^2 \to r_{\mathsf P}^2 (1-{\mathpzc d}_{\mathsf P})$.  The associated pion and kaon impact parameter space GPDs are drawn in Fig.\,\ref{fig:IPDGPDs}.

The contour plots in Fig.\,\ref{fig:IPDGPDs} reveal some interesting features of the three-dimensional distributions of valence constituents within pseudoscalar mesons.
(\emph{a}) All distributions are diffuse at small $|x|$, indicating little probability for finding valence constituents on this domain at $\zeta=\zeta_{\mathpzc H}$.
(\emph{b}) As $|x|$ increases, each probability density acquires a clear maximum at some value of $|b_\perp|$.  The height of the peak increases with $|x|$ whilst its width diminishes.
(\emph{c}) Each distribution has a global maximum; and at the hadron scale, $\zeta_{\mathpzc H}$, their locations and magnitudes (${\mathpzc i}_M$ quoted from Fig.\,\ref{fig:IPDGPDs}) are as follows:
$\pi$ -- $(|x|,b_\perp/r_\pi)=(0.88,0.13)$, ${\mathpzc i}_\pi=3.29$; and
kaon -- $(x,b_\perp/r_K)_{\bar s}=(-0.87,0.13)$, ${\mathpzc i}_K^{\bar s}=3.61$ and $(x,b_\perp/r_K)_u=(0.84,0.17)$, ${\mathpzc i}_K^u=2.38$.
Plainly, within a bound-state seeded by valence constituents with different masses, the heavier object plays a greater role in defining the bound-state's centre of transverse momentum; hence, lies closer to this point.  For the $K$, the relative shift is modest --  just 3\%, whereas the disparity in magnitudes is 20\%, matching $M_{\bar s}/M_u$, which is the typical size for Higgs-modulation of EHM in this system.
It is also worth noting that the pion peak is 10\% higher than the average of the $u$-in-$K$ and $\bar s$-in-$K$ heights.
%
Under QCD evolution, the widths of all peaks broaden as their location drifts toward $x=0$.

Qualitatively and semiquantitatively equivalent impact parameter space GPDs are obtained using the PTIR LFWFs in Sec.\,3 \cite{Raya:2021:progress}.

\smallskip

\noindent\textbf{8.\,Sketching Pion and Kaon Pressure Profiles}.
Consider the first Mellin moment of ${\mathsf P}=\pi, K$ GPDs:
\begin{equation}
\label{Mom1GPD}
%
\int_{-1}^1 \! dx\, x\, H_{\mathsf P}^{\mathpzc q}(x,\xi,-\Delta^2)  = \theta_2^{{\mathsf P}_{\mathpzc q}}(\Delta^2) - \xi^2 \theta_1^{{\mathsf P}_{\mathpzc q}}(\Delta^2)\,,
\end{equation}
${\mathsf P}_{\mathpzc q}=\pi_{u}$, $K_{u}$, $K_{\bar s}$,
where $\theta_{1,2}$ are gravitational form factors: $\theta_1$ is the pressure distribution and $\theta_2$ the mass distribution.

$\theta_2$ is readily computed using the information above, \emph{i.e}.\ the GPDs defined on the DGLAP domain.
$\theta_1$, on the other hand, samples the ERBL region; so, a GPD completion is required.  That may be achieved as follows.
(\emph{i}) Beginning with algebraic PTIRs employed to develop a picture of the pion's GPD \cite{Mezrag:2014jka}, use the Radon transform approach described in Refs.\,\cite{Chouika:2017dhe, Chouika:2017rzs} to complete its extension into the ERBL region, $|x| \leq \xi$.  This approach guarantees that GPD polynomiality and other physical constraints are preserved; and the simple, yet realistic inputs from Ref.\,\cite{Mezrag:2014jka} mean that an algebraic solution of the inverse Radon transform problem is achievable.
(\emph{ii}) The $t$-dependence of the GPD extension obtained thereby is weighted by the dressed-mass of the active valence-parton solution, \emph{e.g}.\ $4 M_u^2$ for the $u$-in-$\pi$ case.  Employing a straightforward generalisation, $u$-in-$K$ involves $4 M_u^2$ and $\bar s$-in-$K$, $4 M_s^2$.  Since these replacements have the status of \emph{Ans\"atze}, we subsequently consider $M \to M (1\pm 0.1)$ in order to estimate a model uncertainty.
(\emph{iii}) Finally, capitalising on explicit calculations and associated insights developed using a symmetry-preserving regularisation of a contact interaction (SCI) \cite{Zhang:2020ecj}, one is led to write
\begin{align}
%
%
& \theta_1^{{\mathsf P}_{\mathpzc q}}(\Delta^2)  = c_1^{{\mathsf P}_\mathpzc q} \theta_2^{{\mathsf P}_{\mathpzc q}}(\Delta^2) \nonumber \\
& + \int_{-1}^1 \! dx\, x\, \left[ H_{\mathsf P}^{\mathpzc q}(x,1,0) P_{M_{\mathpzc q}}(\Delta^2) - H_{\mathsf P}^{\mathpzc q}(x,1,-\Delta^2)\right]\,, \label{EqKaonT1}
\end{align}
where
$c_1^{\pi_u} = 1$,
$c_1^{K_{u,\bar s}}=(1 \pm_{u}^{\bar s} 0.08)$ expresses slight breaking of the soft-pion constraint \cite{Polyakov:1999gs} in the kaon channel, as computed following Ref.\,\cite{Zhang:2020ecj},
and $P_{M}(\Delta^2)= 1/(1+y \ln(1+y))$, $y=\Delta^2/[4 M^2]$, where the $\ln(1+y)$ factor is included to express the form factor scaling violation that is characteristic of quantum field theories in four dimensions.  (Details are provided elsewhere \cite{Raya:2021:progress}.)

Working with Eq.\,\eqref{eq:gaussH}, derived using the factorised \emph{Ansatz} for the $\pi$ and $K$ LFWFs, one obtains the following expressions for the mesons' mass radii, obtained from $\theta_2$ in the same way as the charge radius, $r_{\mathsf P}$, is drawn from $F_{\mathsf P}(\Delta^2)$: 
{\allowdisplaybreaks
\begin{subequations}
\begin{align}
r_{{\mathsf P}_u}^{\theta_2} & = \frac{3 r_{\mathsf P}^2}{2{\mathpzc x}_{\mathsf P}^2} \langle x^2 (1-x) \rangle_{{\mathsf P}_{\bar h}}\,, \\
r_{{\mathsf P}_{\bar h}}^{\theta_2} & = \frac{3 r_{\mathsf P}^2}{2{\mathpzc x}_{\mathsf P}^2}
(1-{\mathpzc d}_{\mathsf P})\langle x^2 (1-x) \rangle_{{\mathsf P}_u}\,,
\end{align}
\end{subequations}}
%
\hspace*{-0.4\parindent}with ${\mathpzc x}_{\mathsf P}^2$ in Eq.\,\eqref{definex}.  Now using Eq.\,\eqref{eq:param} and Table~\ref{tab:DFs}, one obtains
\begin{equation}
\label{kaontheta2radii}
\begin{array}{c|c||c|c}
r_\pi^{\theta_2}/r_\pi & r_{K}^{\theta_2}/r_{K} 
& r_{K_u}^{\theta_2}/{\bar r}_{K} & r_{K_{\bar s}}^{\theta_2}/{\bar r}_{K}\\\hline
0.81 & 0.78 
& 0.84 & 0.72
\end{array}\,,
\end{equation}
where $\bar{r}_K^2 = r_K^2/2$.  The SCI yields \cite{Zhang:2020ecj}: $r_\pi^{\theta_2}/r_\pi=0.89$.
Analogous ratios computed from the $\theta_1$ (pressure) form factors are:
\begin{equation}
\label{kaontheta1radii}
\begin{array}{c|c||c|c}
r_\pi^{\theta_1}/r_\pi & r_K^{\theta_1}/r_K 
& r_{K_u}^{\theta_1}/{\bar r}_{K}
    & r_{K_{\bar s}}^{\theta_1}/{\bar r}_{K}\\\hline
1.18(6) & 1.19(6) 
& 1.25(7) & 1.13(5)
\end{array}\,.
\end{equation}
The SCI produces \cite{Zhang:2020ecj}: $r_\pi^{\theta_1}/r_\pi=1.88(13)$.  Evidently, in all cases treated thus far, the SCI ordering of radii ratios is consistent with our results.
Moreover, this ordering matches that inferred from measurements of $\gamma^\ast \gamma \to \pi^0 \pi^0$ \cite{Kumano:2017lhr}.
Importantly, the separation of baryon number revealed in connection with Eq.\,\eqref{EqmsqLFE} and Fig.\,\ref{fig:IPDGPDs} is also expressed in these gravitational radii.

With results for the $\theta_1$ gravitational form factors in hand, one can compute Breit frame pressure profiles for each associated system \cite{Polyakov:2002yz, Polyakov:2018zvc}.  Using $u$-in-$K$ as an example,
\begin{subequations}
\label{EqPressure}
\begin{align}
p_K^u(r)  & =
 \frac{1}{6\pi^2 r} \int_0^\infty d\Delta \,\frac{\Delta}{2 E(\Delta)} \, \sin(\Delta r) [\Delta^2\theta_1^{K_u}(\Delta^2)] \,, \label{EqPressureA}\\
 s_K^u (r)  & =
%
\frac{3}{8 \pi^2} \int_0^\infty d\Delta \,\frac{\Delta^2}{2 E(\Delta)} \, {\mathpzc j}_2(\Delta r) \, [\Delta^2\theta_1^{K_u}(\Delta^2)] \,, \label{EqPressureB}
\end{align}
\end{subequations}
where
$2E(\Delta)=\sqrt{4 m_K^2+\Delta^2}$
and ${\mathpzc j}_2(z)$ is a spherical Bessel function.
The total pressure sums the contributions from the two valence-parton flavours, \emph{e.g}.\ $p_K = p_K^u + p_K^{\bar s}$.
Here, $p(r)$ is the pressure and $s(r)$ is the shear force.  (Two-dimensional Fourier transforms are sometimes preferred \cite{Miller:2010nz}, as in Eq.\,\eqref{eq:IPDHgen}, but the resulting profiles and magnitudes are similar.)

\begin{figure}[t]
\vspace*{3.5ex}

\leftline{\hspace*{0.5em}{\large{\textsf{A}}}}
\vspace*{-5ex}
\centerline{\includegraphics[clip, width=0.415\textwidth]{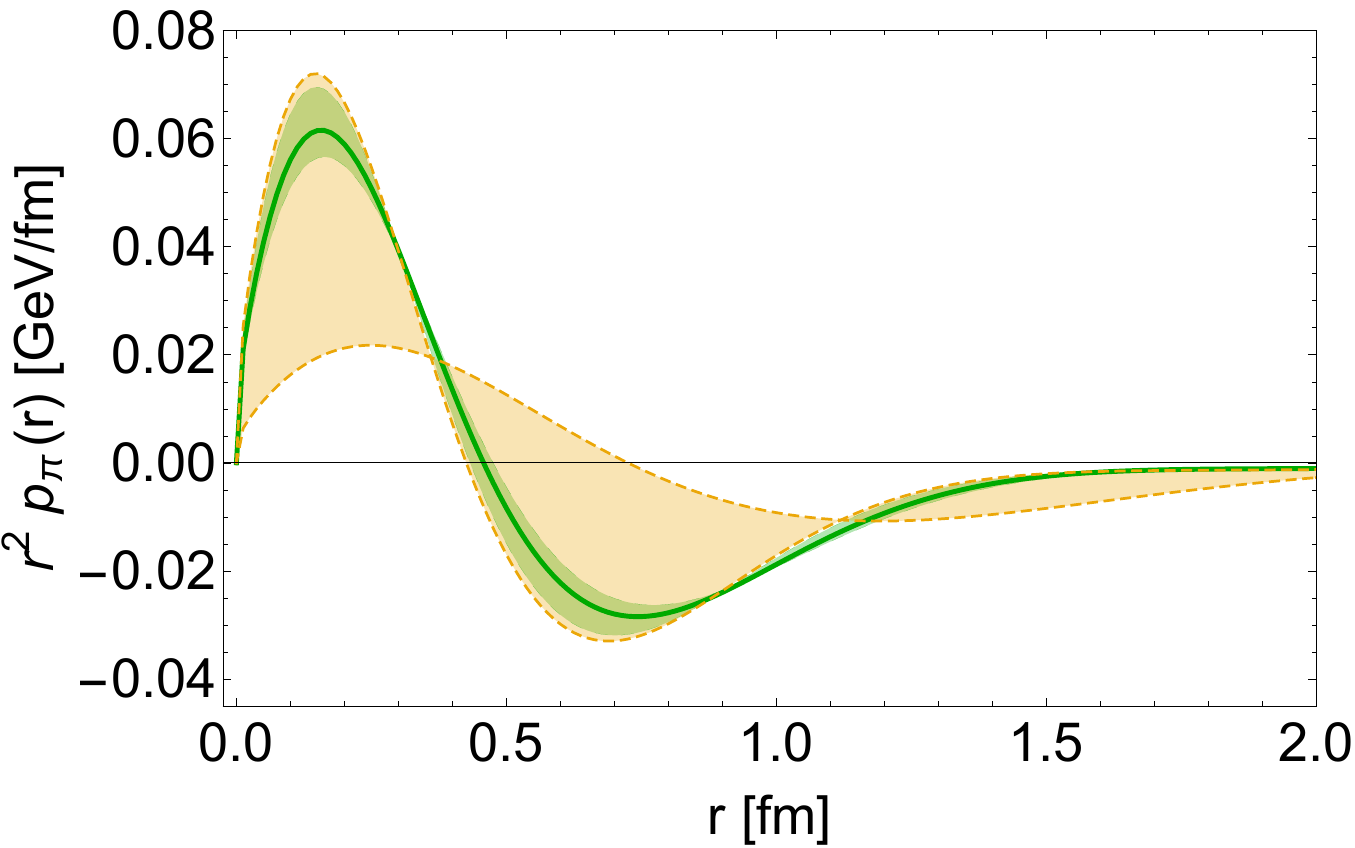}\hspace*{2.2ex}}
\vspace*{6ex}

\leftline{\hspace*{0.5em}{\large{\textsf{B}}}}
\vspace*{-5ex}
\centerline{\includegraphics[clip, width=0.4\textwidth]{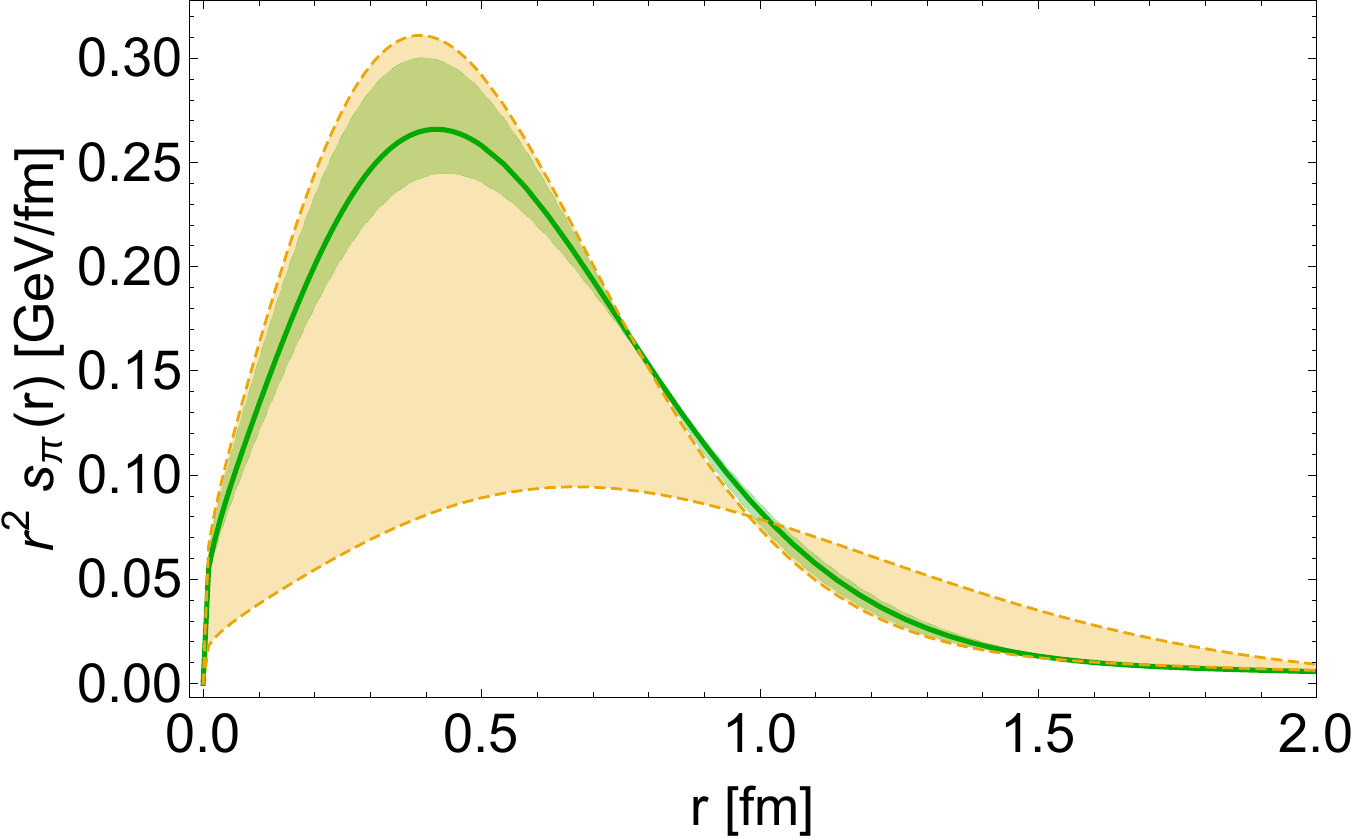}}
%
\caption{\label{fig:r2p} 
\emph{Upper panel}\,--\,{\sf A}. Pion pressure profile.
\emph{Lower panel}\,--\,{\sf B}.  Shear pressure profile.
\textsf{Legend}.
Green curve and band: results computed herein, employing obvious analogues of Eqs.\,\eqref{EqPressure} and Eq.\,\eqref{EqKaonT1} with the terms in the second line evaluated using $M= M_u (1 \pm 0.1)$.
Gold band: range of SCI results \cite{Zhang:2020ecj}.
}
\end{figure}

The pion pressure profiles are depicted in Fig.\,\ref{fig:r2p}.  The qualitative features suggest an intuitive physical interpretation.  Regarding Fig.\,\ref{fig:r2p}A, the pressure is positive and large in the neighbourhood $r\simeq 0$ -- the pion's dressed-valence constituents are pushing away from each other at small separation; but the pressure switches sign as the separation becomes large, indicating a transition to the domain whereupon confinement forces exert their influence on the pair.

The shear pressure, drawn in Fig.\,\ref{fig:r2p}B, is a measure of the strength of forces within the meson which work to deform it.  Evidently, as noted elsewhere \cite{Zhang:2020ecj}, these forces are maximal in the neighbourhood upon which the pressure changes sign, whereat the forces driving the quark and antiquark apart are overwhelmed by attractive confinement pressure.
These perspectives lead one to define a pressure-based confinement radius as the location of the zero in $r^2 p_\pi(r)$:
$r_c^{\pi} = 0.45(3)\,{\rm fm}$,
a result consistent with the SCI value \cite{Roberts:2021ppnp}.

Profiles analogous to Fig.\,\ref{fig:r2p}A can be drawn for neutron stars.  They indicate $r\simeq  0$ pressures therein of roughly $0.1\,$GeV/fm \cite{Ozel:2016oaf}; so, consistent with Ref.\,\cite{Zhang:2020ecj}, we find that the core pressures in $\pi$ mesons and neutron stars are on the same scale.

\begin{figure}[t]
\vspace*{3.5ex}

\leftline{\hspace*{0.5em}{\large{\textsf{A}}}}
\vspace*{-5ex}
\centerline{\includegraphics[clip, width=0.415\textwidth]{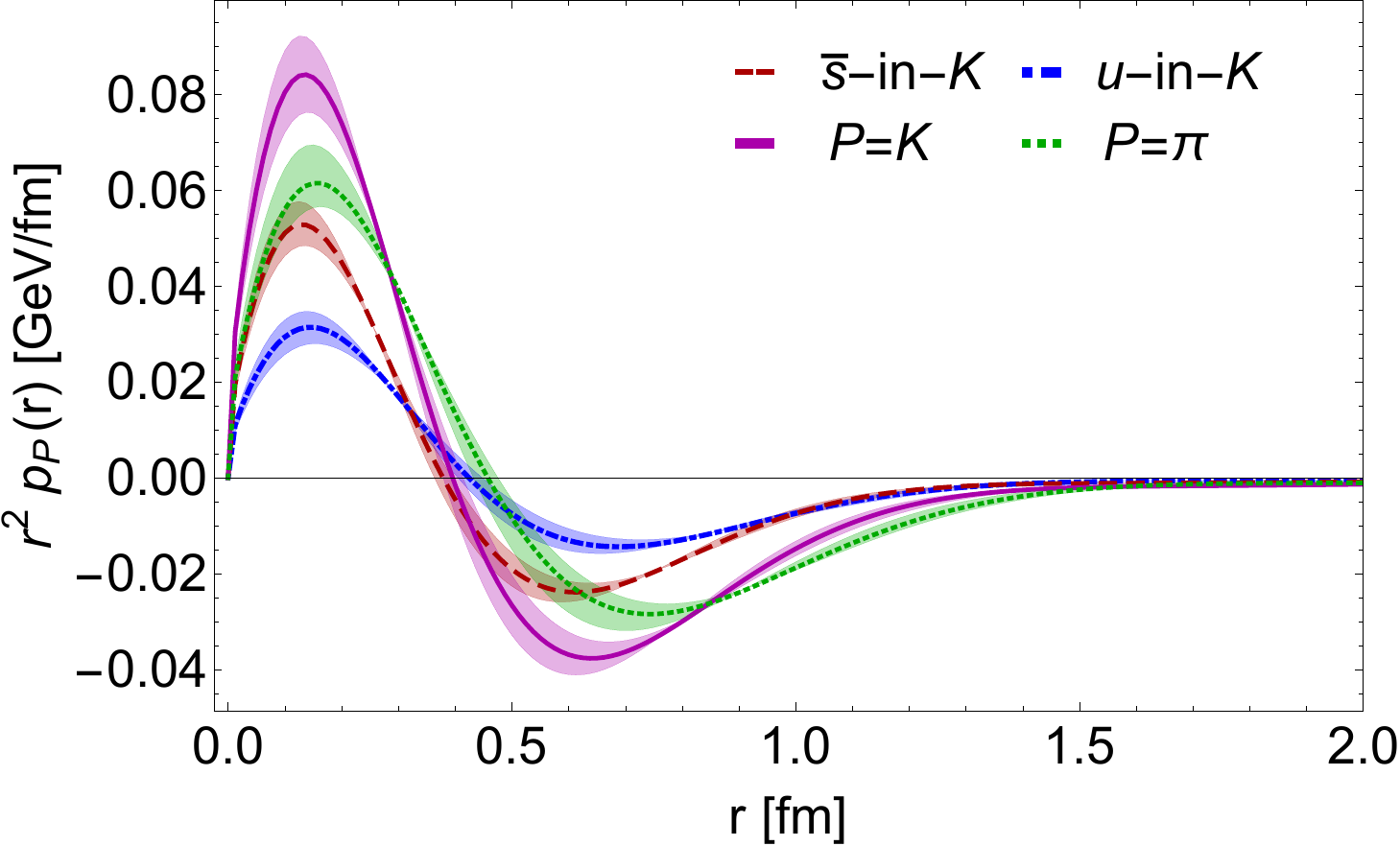}\hspace*{2.2ex}}
\vspace*{6ex}

\leftline{\hspace*{0.5em}{\large{\textsf{B}}}}
\vspace*{-5ex}
\centerline{\includegraphics[clip, width=0.4\textwidth]{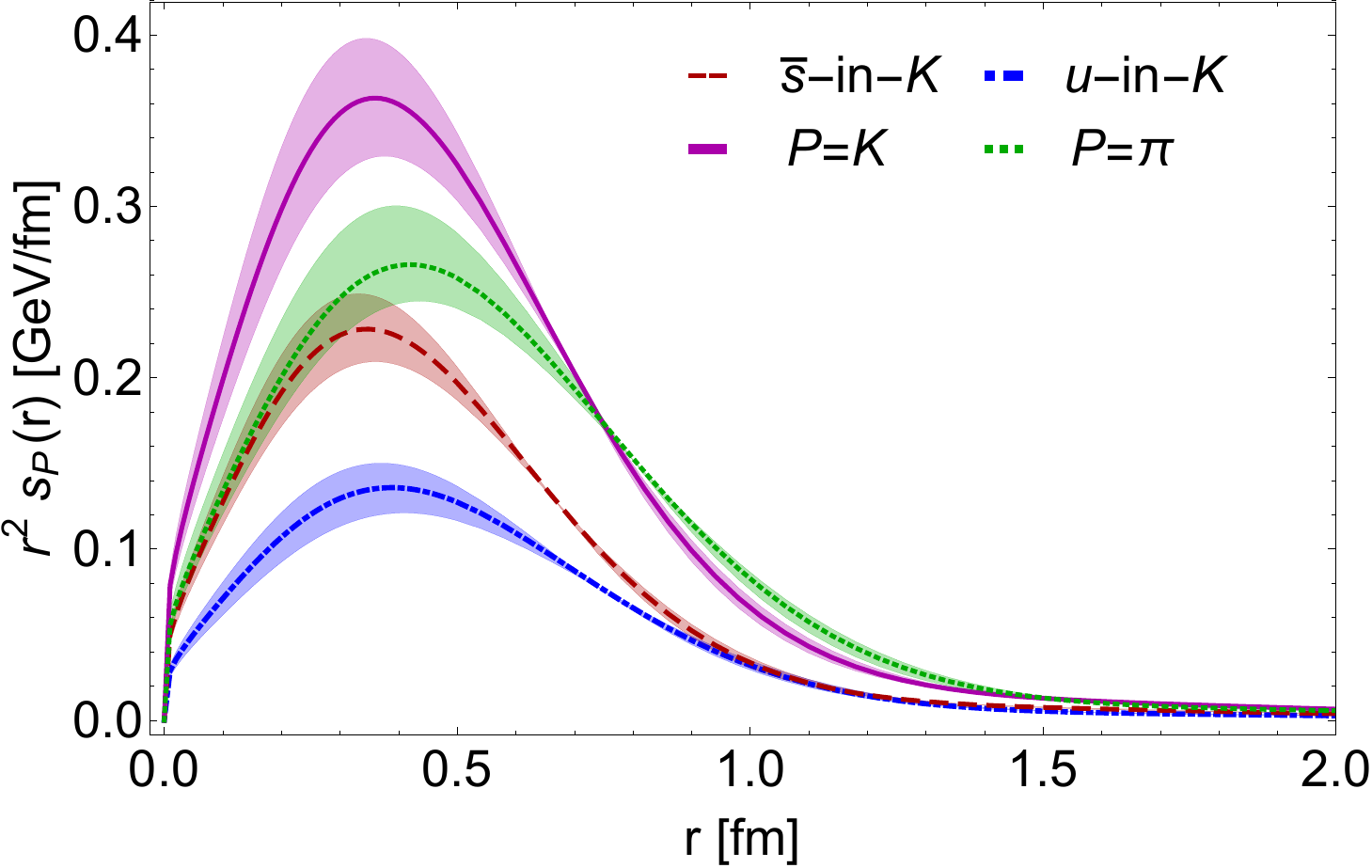}}
%
\caption{\label{fig:r2pK} 
\emph{Upper panel}\,--\,{\sf A}. Pressure profiles in the kaon, Eq.\,\eqref{EqPressureA}.
\emph{Lower panel}\,--\,{\sf B}.  Shear pressure profiles, Eq.\,\eqref{EqPressureB}.
\textsf{Legend}.
Solid magenta curve and band: total $K$ results obtained herein using Eq.\,\eqref{EqKaonT1} with $M_u = M_u(1\pm0.1)$, $M_{\bar s} = M_{\bar s}(1\pm 0.1)$.  Individual $u$-in-$K$ (dot-dashed blue) and $\bar s$-in-$K$ (dashed red) are also displayed.
Green dotted curve and band reproduce Fig.\,\ref{fig:r2p} pion profiles.
}
\end{figure}

The calculated kaon pressures are drawn in Fig.\,\ref{fig:r2pK}.  They are qualitatively identical to those characterising the pion.  Quantitatively, measured by the pressure radius, the $K$ is roughly 15\% smaller than the $\pi$, matching the result inferred from charge radii; and the $K$ core pressure is 20\% larger.  Being bound states, $\int_0^\infty dr\,r^2 p_{\pi,K}(r) = 0$; but, as visually suggested by Fig.\,\ref{fig:r2pK}B:
\begin{equation}
\int_0^\infty dr\,r^2 s_{K}(r) = 1.19(1) \int_0^\infty dr\,r^2 s_{\pi}(r)\,.
\end{equation}

The individual $u$- and $\bar s$-quark contributions to the $K$ pressures are also depicted in Fig.\,\ref{fig:r2pK}.  Consistent with expectations based on
Figs.\,\ref{fig:FFs}, \ref{fig:IPDGPDs}
and
Eqs.\,\eqref{EqmsqLFE}, \eqref{kaontheta2radii}, \eqref{kaontheta1radii}, our analysis predicts that the
kaon's $\bar s$-quark contributes a greater fraction of the total $K$ pressure than its partner $u$-quark,
its peak/trough intensities are greater,
and the associated distributions are localised nearer to $r=0$.
These observations and the presence of strong scalar and axial-vector diquark correlations in the proton \cite{Barabanov:2020jvn} suggest that one may expect similar effects in the flavour separation of the proton's pressure profiles, with $u_p$ acting like $\bar s_K$ and $d_p$ like $u_K$.  There are hints of this in Ref.\,\cite[Fig.\,8]{Cui:2020rmu}.

\medskip
\noindent\textbf{9.\,Summary and Perspective}.
Beginning with pion and kaon valence distribution functions that explain available data, we developed algebraic factorised \emph{Ans\"atze} for the associated light-front wave functions (LFWFs) and, therefrom, $\pi$ and $K$ generalised parton distributions (GPDs) on the DGLAP domain.  The GPDs were validated by comparing the derived meson elastic electromagnetic form factors, $F_{\pi,K}$, with modern data.   Subsequently calculating the impact parameter space GPDs, a separation of baryon number was found within the kaon, with the $\bar s$ quark being localised closer to the kaon's centre of transverse momentum than the $u$ quark.

An extension of the GPDs onto the ERBL domain was accomplished using a Radon transform approach applicable to algebraic LFWFs, enabling calculation of the $\pi$ and $K$ gravitational form factors, $\theta_{1,2}$.  In each channel ${\mathsf P}=\pi, K$, as a function of $\Delta^2>0$, $\theta_2^{\mathsf P}$ is harder/flatter than $F_{\mathsf P}$; in turn, $F_{\mathsf P}$ is harder than $\theta_1^{\mathsf P}$.  Furthermore, each $K$ form factor is harder than its $\pi$ analogue; and $\bar s$-in-$K$ form factors are harder than kindred $u$-in-$K$ functions.

Working with Breit-frame pressure and shear force profiles defined in terms of $\theta_1^{\mathsf P}$, we found that
$K$ profiles are more compact than $\pi$ profiles and both achieve near-core pressures of similar magnitude to that found in neutron stars;
the total shear pressure in the $K$ exceeds that in the $\pi$;
and $\bar s$ profiles within the kaon are more compact than the partner $u$ profiles.
The size of all such differences is typical of Higgs-boson induced modulations of the basic emergent hadron mass scale.

Natural extensions of this study include
the use of unfactorised, perturbation theory integral representations of the LFWFs in the calculation of all quantities considered herein -- an effort which is progressing well;
a direct extension of the derived DGLAP GPDs onto the ERBL domain using a numerical approach to the Radon transform problem -- a project that is underway;
and development of analogous paths to the construction of realistic nucleon GPDs, which is an open challenge.

\medskip
\noindent\emph{Acknowledgments}.
We are grateful for constructive comments from
F.~De~Soto, C.~Mezrag, H.~Moutarde, J.-L.~Ping and J.~Segovia.
%
Work supported by:
National Natural Science Foundation of China (grant 11805097);
Jiangsu Provincial Natural Science Foundation of China (grant BK20180323);
University of Huelva (grant EPIT-2019);
Jiangsu Province \emph{Hundred Talents Plan for Professionals};
Spanish Ministry of Science and Innovation (MICINN) (grant PID2019-107844GB-C22);
and
Junta de Andaluc\'ia (contracts P18-FRJ-1132, P18-FR-5057).
%
%



\end{document}